\documentclass[fleqn,usenatbib]{mnras}

\usepackage{newtxtext,newtxmath}

\usepackage[T1]{fontenc}
\usepackage[textsize=tiny, backgroundcolor=orange!50]{todonotes}

\DeclareRobustCommand{\VAN}[3]{#2}
\let\VANthebibliography\thebibliography
\def\thebibliography{\DeclareRobustCommand{\VAN}[3]{##3}\VANthebibliography}

\usepackage{mathrsfs}  
\usepackage{graphicx} 
\usepackage{amsmath}  
\usepackage{mathtools}
\usepackage{bm}

\usepackage{tikz}
  \usetikzlibrary{patterns, angles, quotes}
  \usetikzlibrary{decorations.pathreplacing}

  \pgfdeclarepatternformonly[\hatchspread,\hatchthickness,\hatchshift,\hatchcolor]
     {custom north east lines}
     {\pgfqpoint{\dimexpr-2\hatchthickness}{\dimexpr-2\hatchthickness}}
     {\pgfqpoint{\dimexpr\hatchspread+2\hatchthickness}{\dimexpr\hatchspread+2\hatchthickness}}
     {\pgfqpoint{\dimexpr\hatchspread}{\dimexpr\hatchspread}}
     {
      \pgfsetlinewidth{\hatchthickness}
      \pgfpathmoveto{\pgfqpoint{\dimexpr\hatchshift-0.15pt}{-0.15pt}}
      \pgfpathlineto{\pgfqpoint{\dimexpr\hatchspread+0.15pt}{\dimexpr\hatchspread-\hatchshift+0.15pt}}
      \ifdim \hatchshift > 0pt
        \pgfpathmoveto{\pgfqpoint{-0.15pt}{\dimexpr\hatchspread-\hatchshift-0.15pt}}
        \pgfpathlineto{\pgfqpoint{\dimexpr\hatchshift+0.15pt}{\dimexpr\hatchspread+0.15pt}}
      \fi
      \pgfsetstrokecolor{\hatchcolor}
      \pgfusepath{stroke}
     }
\usepackage{tikzscale}
\usepackage{tkz-euclide}

\usepackage{multicol}
\usepackage{graphicx}

\newcommand{\GAMMA}{\texttt{GAMMA}}

\title[The GAMMA code]{GAMMA: a new method for modeling relativistic hydrodynamics and non-thermal emission on a moving mesh}

\author[E. H. Ayache et al.]{
Eliot H. Ayache,$^{1}$\thanks{E-mail: e.h.r.ayache@bath.ac.uk (EHA)}
Hendrik J. van Eerten,$^{1}$\thanks{E-mail: h.j.van.eerten@bath.ac.uk (HJVE)}
Rupert W. Eardley $^{1}$
\\
$^{1}$Department of Physics, University of Bath, Claverton Down, BA2 7AY, UK\\
}

\date{Accepted XXX. Received YYY; in original form ZZZ}

\pubyear{2020}

\begin{document}
\label{firstpage}
\pagerange{\pageref{firstpage}--\pageref{lastpage}}
\maketitle

\begin{abstract}

  In recent years, dynamical relativistic jet simulation techniques have progressed to a point where it is becoming possible to fully numerically resolve gamma-ray burst (GRB) blast-wave evolution across scales. However, the modeling of emission is currently lagging behind and limits our efforts to fully interpret the physics of GRBs. In this work we combine recent developments in moving-mesh relativistic dynamics with a local treatment of non-thermal emission in a new code: \GAMMA. The code involves an arbitrary Lagrangian-Eulerian approach only in the dominant direction of fluid motion which avoids mesh entanglement and associated computational costs. Shock detection, particle injection and local calculation of their evolution including radiative cooling are done at runtime. Even though \GAMMA~has been designed with GRB physics applications in mind, it is modular such that new solvers and geometries can be implemented easily with a wide range of potential applications. In this paper, we demonstrate the validity of our approach and compute accurate broadband GRB afterglow radiation from early to late times. Our results show that the spectral cooling break shifts by a factor of $\sim40$ compared to existing methods. Its temporal behavior also significantly changes from the previously calculated temporary steep increase after the jet break. Instead, we find that the cooling break does not shift with time between the relativistic and Newtonian asymptotes when computed from our local algorithm.
  \GAMMA~is publicly available at: \url{https://github.com/eliotayache/GAMMA}.

\end{abstract}




\section{Introduction}


  The simulation of gamma-ray-burst (GRB) afterglow evolution is a particularly challenging multiscale numerical problem \citep{Granot2007,VanEerten2017}. These collimated relativistic jets produced by the collapse of a massive star (long GRBs) \citep{Woosley1993,MacFadyen1999a} or a compact binary merger (short GRBs) \citep{Eichler1989,1995A&A...293..803M} exhibit features crucial to our understanding of their behaviour over several orders of magnitude in time and space. Various analytical and semi-analytical models for the lateral spreading of afterglow jets  based on single shell models exist in the literature \citep{Rhoads1999b,VanEerten2010a,Wygoda2011,VanEerten2012b,Granot2012,VanEerten2013b,Duffell2017,Ryan2019a}. However, even when calibrated on simulations these do not capture the full radial and angular profiles computed by simulations. In the last 20 years, dynamical simulations of GRB blast waves have evolved from one-dimensional (1D) Lagrangian computations of the evolution of spherically symmetric fireballs \citep{Kobayashi1999,Daigne2000}, to state-of-the-art two-dimensional (2D) and three-dimensional (3D) Eulerian simulations \citep{Kumar2003,Cannizzo2004,Zhang2009,VanEerten2010a,Meliani2010,DeColle2011,Wygoda2011,VanEerten2012}. The latter can simulate the sideways interaction of collimated jets with the circumburst (CSM) medium and provide insight in the stability of the ejecta-CSM interface. Unfortunately, these approaches remain particularly computationally expensive. They rely on intense adaptive mesh refinement (AMR) procedures in order to capture the extreme resolution needed to properly resolve the head of the jet and converge before the onset of the jet's sideways expansion. As it is notoriously difficult to resolve jet spreading behavior in the lab frame even with AMR, a first successful approach improving convergence and computational efficiency has been to move the computation in a Lorentz-boosted frame \citep{VanEerten2013a}. Moving at fixed velocity along the jet axis, this frame minimises the Lorentz-contraction of the blast-wave and relaxes the resolution constraints.
 
  The use of arbitrary Langrangian-Eulerian (ALE ) methods helps to improve the numerical resolution of simulations of astrophysical flow, as has been demonstrated for Newtonian dynamics by e.g. AREPO \citep{Springel2010,Weinberger2019}. In this finite-volume approach, the mesh edges can be moved arbitrarily during the dynamical evolution. In practice, matching the mesh motion to that of the fluid provides significant improvement in terms of time-stepping and resolution around shocks as the numerical prescription effectively becomes pseudo-Lagrangian. \citet{Duffell2011} implemented this approach in the special relativistic context in their code TESS, finally opening the door to numerically capturing the trans-relativistic evolution of GRB jets across scales from the ultra-relativistic stage to the deceleration. Furthermore, in cases in which fluid motion is dominant in one direction, further progress was made by sidestepping the need for computationally expensive re-gridding operations usually associated with moving meshes. JET \citep{Duffell2013} and DISCO \citep{Duffell2016} take advantage of this directionality and model the dynamics on parallel 'tracks' along which fluid zones can move freely, leading to a significant increase in computational efficiency. 

  As a result, the bottleneck for accurate modeling of highly energetic transients currently lies in the calculation of associated radiative emission. GRB afterglows are the result of synchrotron emission from shocks forming in the head of the jet as it interacts with the CSM \citep{Rees1992,Meszaros1997}. Current numerical radiative prescriptions rely on approximations of the evolution of the micro-physical state in the fluid downstream of these shocks. The widely used global cooling approximation assumes a single micro-physical state across the whole fluid profile that evolves globally with time since the explosion \citep{Sari1997a}. While this conserves the scalings and closure relations in each regime of the resulting spectra, this approximation is known to produce errors in the absolute flux level by up to a decade, as well as an incorrect position of the characteristic spectral break frequencies \citep{VanEerten2010a,Guidorzi2014}. This makes any broadband interpretation of the data across timescales very difficult. While analytical solutions locally calculating the micro-physical states have been used for 20 year now \citep{Granot2001} and produce satisfying results in the ultra-relativistic limit of top-hat jets observed on-axis, an accurate description across all stages of jet evolution is still missing. Having at our disposal a numerical tool capable of computing such radiation accurately, efficiently and with precision promises to finally allow broadband fitting of afterglow light-curves and spectra. This toolkit could achieve this by refining the current templates and providing benchmarking opportunities for more efficient semi-analytical approaches, while also allowing for the accurate study of edge cases involving complex dynamics and multiple radiation emission sites.

  In this work, we present a new numerical code, \GAMMA, and use it to show the striking difference obtained in the radiative evolution in the trans-relativistic phase of the jet's life compared to previous approaches. In order to achieve this calculation, \GAMMA~combines the moving mesh approach to multi-dimensional dynamical simulations seen in JET and DISCO with a local calculation of the micro-physical accelerated particle population evolution. The local cooling approach is possible thanks to the increased resolution from the moving mesh around the shocks that allows accurate computation of the rapidly evolving particle energy distribution. In section \ref{sec:code_description} we describe the dynamical part of the code. Section \ref{sec:tests} is dedicated to standard tests of the dynamics. We also investigate the code's ability to capture complex dynamics by reproducing results from a study of Rayleigh-Taylor (RT) instabilities at the contact discontinuity between ejecta and CSM \citep{Duffell2013}. We then describe the local cooling prescription in section \ref{sub:local_synchrotron_cooling}. Finally, in section \ref{sub:synthetic_grb_afterglow_light_curves_with_local_cooling_from_2d_simulations} we present the calculation of accurate synthetic afterglow light-curves and spectra from early to late times from the forward shock (FS) from 2D axisymmetric simulations of a GRB jet. A discussion of the implications and limitations of our work is presented in in section \ref{sec:discussion}.

\section{Code description}
  \label{sec:code_description}

  \GAMMA~ uses a Godunov scheme to solve hyperbolic systems of partial differential equations (PDEs) in one and two dimensions (3-dimensional evolution will be implemented at a later stage). The solvers currently implemented correspond to special relativistic hydrodynamics (SRHD). New solvers (e.g magneto-hydrodynamics) can be added easily. The code follows the same approach as JET and DISCO \citep{Duffell2013,Duffell2016} with the addition of the radiation module and a treatment of parallelisation that takes advantage of shared memory architectures. In this section we describe the numerical approach to the dynamics. The radiative local particle acceleration and cooling and the associated radiation are described in section \ref{sub:local_synchrotron_cooling}.

  \subsection{Special relativistic hydrodynamics} 
  \label{sub:special_relativistic_hydrodynamics}
  
    The fluid can be described by a state vector of \emph{primitive variables} $\bm{V} = (\rho, \vec{v}, p)^T$, where $\rho$ and $p$ are the rest-mass density and pressure in the co-moving frame, and $\vec{v}$ is the fluid velocity in the lab frame. We solve the following system of equations:
    \begin{align}
      \partial_t \bm{U} + \bm{\nabla F}(\bm{U}) = \bm{S},
      \label{eq:conservativeFormulation}
    \end{align}

    Where $\bm{U}$ and $\bm{F}(\bm{U})$ are the vector of \emph{conserved variables} and the corresponding flux vector, respectively, and $\bm{S}$ is the source term. $\bm{\nabla}$ is the divergence operator broadcast on all spatial vector components of $\bm{F}(\bm{U})$. $\bm{U}$ and $\bm{F}(\bm{U})$ are expressed in terms of primitive variables as follows:

    \begin{align}
      \bm{U} =
      \begin{pmatrix}
        D \\ 
        \vec{m} \\
        \tau 
      \end{pmatrix}
      \equiv
      \begin{pmatrix}
        \rho \Gamma \\ 
        \rho h \Gamma^2 \vec{v}\\ 
        \rho h \Gamma^2 - p - D
      \end{pmatrix}  &&
      \begin{matrix*}[l]
        \text{(Rest-mass density)}\\
        \text{(Momentum)}\\
        \text{(Energy)}
      \end{matrix*},
    \end{align}
    \begin{align}
        \bm{F}_i(\bm{U}) = 
        \begin{pmatrix}
          D v_i \\ \vec{m} v_i + p \hat{i} \\ m_i - D v_i,
        \end{pmatrix}, && \forall i \in \{x,y,z\},
    \label{eq:conserved_variables}
    \end{align}

    where $\hat{i}$ is the unit vector in the i-direction, $h$ is the specific enthalpy including rest-mass energy in the co-moving frame, $\Gamma$ is the Lorentz factor, and the speed of light is set to $c=1$. The SRHD equations can be cast in their angular momentum conserving form identical to eq.~\ref{eq:conservativeFormulation} for cylindrical $(r, \theta, z)$ and spherical $(r, \theta, \phi)$ coordinates \citep{Mignone2007}. This requires that we replace, in the conservation equation, the $\theta$ component of linear momentum $m_\theta$ in $\bm{U}$ with the the angular momentum $r m_\theta$, and the flux of the $\theta$ momentum $F_{i\theta} = m_\theta v_i + p \delta^{i}_\theta$ with $rF_{i\theta}$. With this form of the equations, the following source terms appear in 2D:
    \begin{align}
      \text{cylindrical} && \bm{S} = (0~,  p/r~, 0~, 0)^T ,\\
      \text{spherical}   && \bm{S} = (0~, (\rho h \Gamma^2 v_\theta^2 + 2p )/r, p / \tan \theta~, 0)^T,
    \end{align} 
    where the pressure terms compensate our inclusion of $p$ in the divergence and the other term is a "geometrical" source term.
    The calculation of the geometrical source terms for the linear momentum conserving form, for any set of orthogonal curvilinear coordinates, is presented in the appendix of \citet{Mignone2005}. The corresponding derivation in the case of SRHD is reported in the appendix of \citet{Zhang2006}, which is equivalent to our approach. The full conservation equations can also be derived in any curved metric using the "Valencia formulation" \citep{Banyuls1997} against which we compared our expressions.

    We close the system of equations with the Synge-like ideal mono-atomic fluid equation of state (EOS) from \citet{Meliani2004a} based on the relativistic perfect gas law \citep{Synge1957a,Mathews1971a}:
    \begin{align}
      p(\rho, \epsilon) = \rho \epsilon(\gamma_\mathrm{eff} - 1),
    \end{align}
    where $\epsilon$ is the specific internal energy density and $\gamma_\mathrm{eff}$ the effective polytropic index of the fluid given by:
    \begin{align}
      \gamma_\mathrm{eff} = \gamma - \frac{\gamma-1}{2} \left( 1 - \frac{1}{e^2} \right).
    \end{align}
    $\gamma = 5/3$ is the fixed adiabatic index of the fluid in the non-relativistic (cold) case and $e$ the specific internal energy including rest-mass.
    $\gamma_\mathrm{eff}$ is dependent on the fluid temperature such that $\gamma_\mathrm{eff} = 4/3$ in the ultra-relativistic case and $\gamma_\mathrm{eff} = 5/3$ in a non-relativistic fluid and allows for a trans-relativistic description of the evolution. This description is a very good approximation to the Synge gas equation and avoids the costly evaluation of associated Bessel functions.


  \subsection{Riemann solver} 
  \label{sub:riemann_solver}

    \begin{figure}
      \centering
      \includegraphics[width=0.45\textwidth]{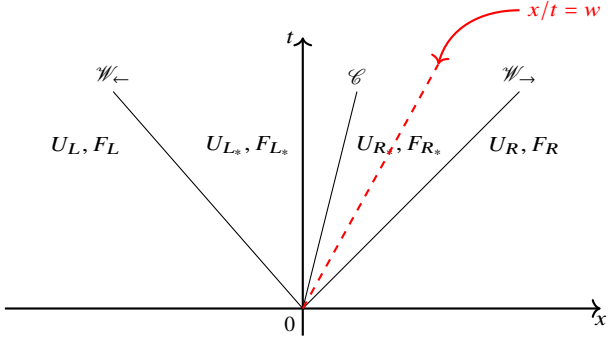}
      \caption{Schematic representation of a Riemann problem. The dashed red line shows the movement of the interface at velocity $w$. The flux across the interface will, and the state chosen to correct for the motion of the interface will be $\bm{F}_{R_*}$ and $\bm{U}_{R_*}$, respectively.}
      \label{fig:Riemann_problem}
    \end{figure}
  
    The code is based on a finite-volume Godunov scheme. The simulation domain is divided in discrete volumes, or cells, in which the local fluid state is averaged. To evolve the system we solve a Riemann problem at each interface by calculating the corresponding Riemann fan of waves emerging from the discontinuity and the associated fluxes. At this stage, \GAMMA~ includes the HLLC \citep{Mignone&Bodo2006} solver for relativistic hydrodynamics. This solver builds on the two-wave HLL solver \citep{Harten1983} by adding a calculation of the wavespeed of the contact discontinuity (CD). As explained in the next section, we set the interface velocity to that of the CD and thus require the use of a complete three-wave solver.

    \GAMMA~ follows an arbitrary Lagrangian-Eulerian approach (ALE). This means that it can compute inter-cell fluxes for arbitrary interface velocities, in any direction. Figure \ref{fig:Riemann_problem} describes a Riemann problem for a moving interface with velocity $w$. For the HLLC hydrodynamics solver, three waves ($\mathcal{W}_\leftarrow$, $\mathcal{C}$, $\mathcal{W}_\rightarrow$) emerging from the discontinuity split the fluid in 4 regions ($L$, $L_*$, $R_*$, $R$). The flux across the interface is given by $\bm{F} = \bm{F}_\mathrm{Riemann} - w \bm{U}_\mathrm{Riemann}$, where $\bm{F}_\mathrm{Riemann}$ and $\bm{U}_\mathrm{Riemann}$ are the flux and state vectors of the fluid in the region in which sits this interface (region $R_*$ in the situation depicted in figure \ref{fig:Riemann_problem}).


  \subsection{Moving mesh and parallelisation} 
  \label{sub:moving_mesh}

  \begin{figure}
    \centering
    \includegraphics[width=0.5\textwidth]{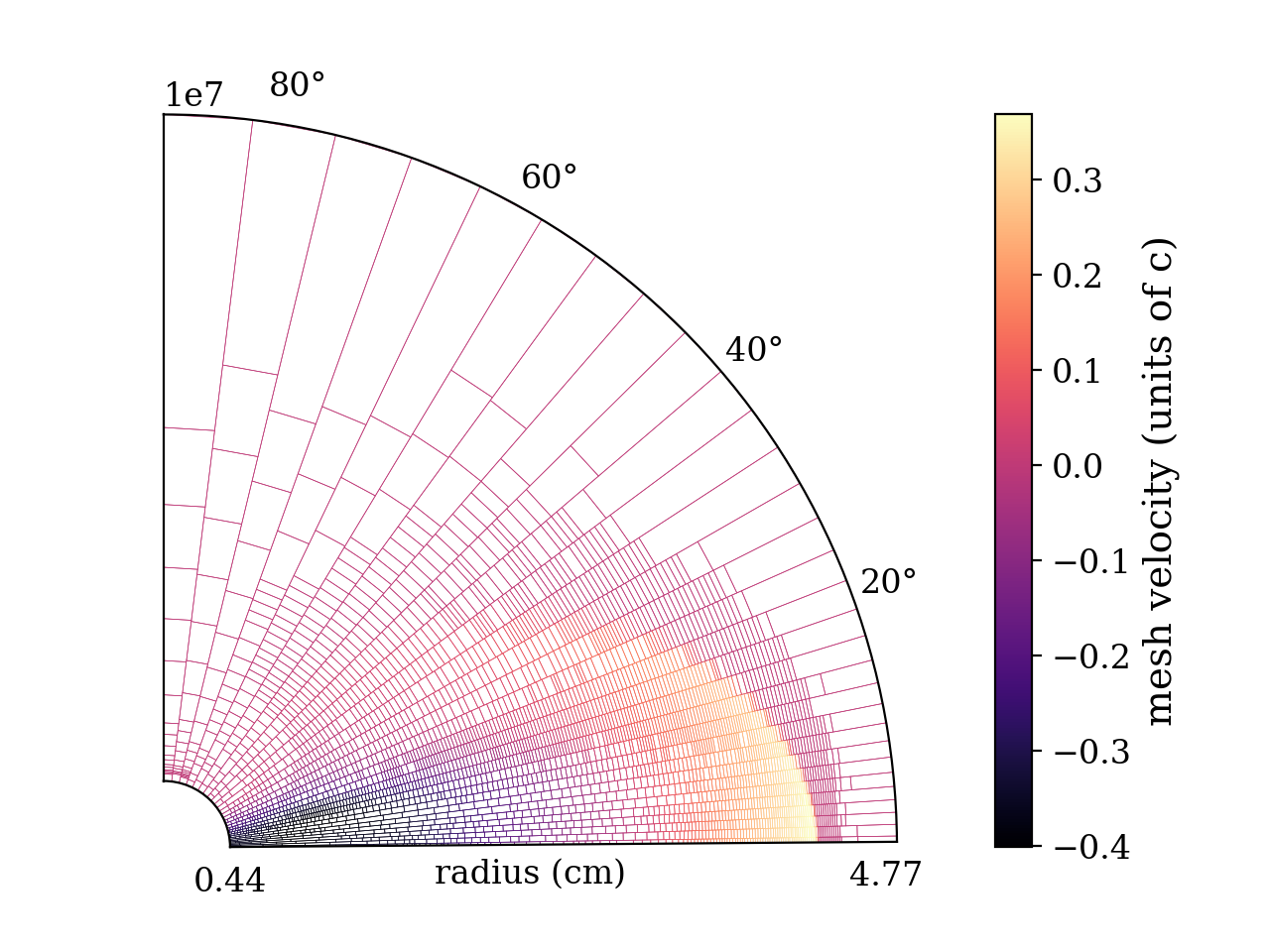}
    \caption{Mesh radial velocity and interface positions for a low-resolution example of a relativistic GRB jet. Axi-symmetry at $\theta=0$ and planar symmetry at $\theta=90\deg$ allow us to restrict the simulation to $\theta \in [0,\pi/2]$. The resolution naturally increases at the shock without the need for active adaptive mesh refinement.}
    \label{fig:tracks}
  \end{figure}

    Moving meshes provide several advantages to simulate the evolution of GRB afterglows, but more broadly to simulate dynamics over a wide range of scales. First, the time-step can be increased for smooth regions of high velocity, where it is essentially limited by the speed of sound, in comparison to fixed-mesh approaches where the bulk velocity is the limiting factor. This is of particular interest to us as we look to maximising the resolution downstream of shocks to capture local cooling accurately. Second, a moving mesh matching the fluid velocity ends up naturally refining the regions of strong gradients as the fluid zones pile-up in compression waves and shocks. Because we are trying to capture the very fast evolving particle population downstream of shocks, a pseudo-Lagrangian approach is ideal.

    The mesh is allowed to move in one direction, which we will assume to be the $x$-direction for the rest of this section. The transverse direction will be called $y$ as the treatments remain the same in 2D or 3D. The simulation grid is built as a set of tracks along which the interfaces between the cells are allowed to move at arbitrary velocity. This frees us from the re-gridding operations associated with mesh entanglement, and ensures that interfaces remain orthogonal to the coordinate system basis vectors. In practice, to maximise mass conservation in a given cell, we set interface velocity to the contact discontinuity velocity ($\mathcal{C}$ wave in the Riemann fan). Figure \ref{fig:tracks} illustrates how the mesh moves following these tracks. Allowing the mesh to move forces us to re-compute the geometry of the interfaces between tracks for each time-step. We do this by looping over all the couples of neighbouring tracks each time instead of keeping track of these interfaces from one time-step to another.

    We use a hybrid OpenMP/MPI approach to parallelisation in order to make use of shared memory on a single node. This gives us more flexibility when dealing with variable numbers of cells per track. The simulation domain is cut in the $y$ direction into a number $N_\mathrm{nodes}$ of regions containing an equal number of neighboring tracks, each region sent to a single node. Ultimately, depending on the number of nodes, tracks per node, and cells per track, the user will be able to choose to parallelise the computation in each region using OpenMP either in the moving $x$ direction (for a given track, cell are distributed over the available cores) or in the fixed $y$ direction (for a given region, track are distributed over the available cores). Parallelising in $x$ minimises the number of cores in an idle state in a node, but runs into the risk of false sharing for tracks with very few cells compared to the number of cores available, which is why we first implemented the parallelisation in $y$ with the $x$ version available soon. One limitation to our approach is when the average number of cells per track varies strongly from region to region (typical case in figure \ref{fig:tracks}), which leads to some nodes having much fewer cells to evolve and sit idle. At this stage we compensate this by giving more tracks to these nodes to compute. Further improvements on parallelisation are expected in future versions of \GAMMA.


  \subsection{Spatial reconstruction} 
  \label{sub:spatial_reconstruction}

    \begin{figure}
      \centering
      \includegraphics[width=0.45\textwidth]{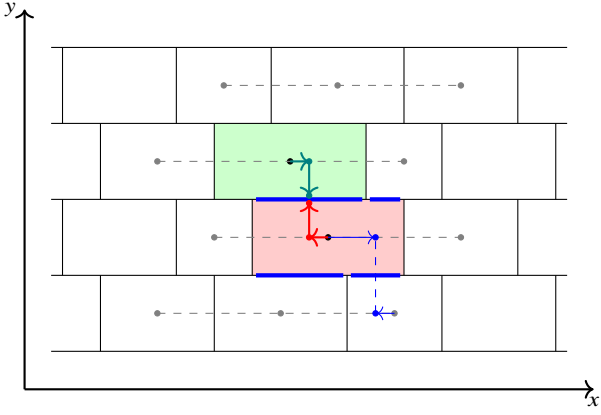}
      \caption{Schematic representation of the procedure to reconstruct the states left and right of the interface between the red and green cells. For example for the red cell, the first step is to compute the transverse gradients for all the $y$-interfaces of the cell (marked in blue). These gradients are measured between states projected on the $x$ coordinate corresponding the center of the interface using the $x$ gradients already computed (gray dashed lines) We then compute the average gradient for the cell and apply the slope limiter. Using this gradient for each cell, it's possible to reconstruct the red and green states.}
      \label{fig:spatial_reconstruction}
    \end{figure}

    \GAMMA~ currently includes piecewise linear spatial reconstruction. Figure \ref{fig:spatial_reconstruction} shows a schematic representation of gradient calculation in two  dimensions. In the $x$-direction, the reconstruction can be done independently in each track as the cell centers and the center of the interface between cells are aligned. we store the slope-limited gradients in the $x$-direction for use in the $y$-direction gradient calculation. Indeed, for reconstruction across tracks, we follow the procedure from \citet{Duffell2016}. The gradient $\vec{\bm{w}}$ inside a given cell is calculated using the following steps.
    First, for every interface $i$ between two cells at $(x_i, y_i)$ coordinate, we compute an associated gradient in the $y$-direction $\bm{w}_{y,i}$ across this interface:
    \begin{align}
      &\bm{w}_{y,i} = [\bm{W}^+(x_i) - \bm{W}^-(x_i)] / (y^+ - y^-) \\
      &\text{with } \bm{W}^\pm(x) = \bm{W}^\pm_0 + \bm{w}^\pm_x (x - x^\pm_0)
    \end{align} 
    where the $+$ and $-$ exponents denote the cell above and below the interface, respectively (see figure \ref{fig:spatial_reconstruction}), $\bm{W}_0^\pm$ is the cell-centered primitive fluid state and $\bm{w}_x^\pm$ the gradient in the $x$-direction.
    We then compute an averaged gradient $\bar{\bm{w}}_y$ for the considered cell from the gradients associated with all the interfaces composing its two y-faces, weighted by their respective surface areas $A_{y,i}$:
    \begin{align}
      \bar{\bm{w}}_y = \frac{\sum_{i \in \{\mathrm{interfaces}\}} \bm{w}_{y,i}A_i}
      {\sum_{i \in \{\mathrm{interfaces}\}} A_i}
    \end{align}
    To avoid spurious oscillations we apply a minmod slope limiter such that the final gradient value in the $y$-direction is:
    \begin{align}
      \bm{w}_y = \mathrm{minmod}(\bar{\bm{w}}_y, (\bm{w}_{y,i})_{i\in{\{\mathrm{interfaces}\}}}),
    \end{align}
    Once the gradients have been computed in all directions for every cell, it is then possible to  finally compute the fluid states on either sides of every interface by interpolating from the cells centers:
    \begin{align}
      \bm{W}(\vec{r}_i) = \bm{W}_0 + \vec{r}_i \cdot \vec{\bm{w}},
    \end{align}
    where $\vec{r}_i$ is the interface coordinate calculated from the cell centroid to ensure quantity conservation during spatial reconstruction.


  \subsection{Time-stepping and adaptive mesh refinement} 
  \label{sub:adaptive_mesh_refinement}
  
    The cell conserved quantities are updated by summing over all fluxes across their associated interfaces in all directions, for each time-step (method of lines), while accounting for potential source terms. The moving interface positions are updated according to their measured velocities during this time-step too, which leads to minimum fluxes (and zero flux in mass) across them when setting their velocity to that of the CD.
    The time-integration can be chosen between Euler time-stepping and third order Runge-Kutta, which has the advantage of being total-variation-diminishing.
    We use an adaptive time-step based on a Courant-Friedrich-Lewy (CFL) condition \citep{Courant1928}. This introduces issues when combined with a moving mesh as compressed fluid cells will lead to a detrimental decrease in time-step if no lower bound is chosen for their size. We implement the ability for the user to set up their own criteria for adaptive mesh refinement and include in the code methods to split and merge cells together on a single track. The code offers two different modes for re-gridding: a "runaway" mode in which the total number of cells on a given track is only constrained by a maximum value, and a "circular" mode in which every call of the merge/split function calls an instance of the split/merge function on a cell in the same track based on a calculation of its "re-gridding score" that can also be set by the user. This "circular" mode ensures that the total number of cells on a single track is constant throughout the simulation.

    The new physical state in cells post-merger are is averaged over the two states in the cells prior to merger, ensuring quantity conservation. Cell-splitting follows linear interpolation of the conserved variables from the new cells centroids. We apply the same slope limiter as in the spatial reconstruction on the gradient used for the interpolation. This ensures quantity conservation and limits oscillations around shock fronts.


  \section{Tests} 
  \label{sec:tests}

    We evaluate the accuracy and convergence of the code on a range of standard tests. All tests are carried out with a CFL condition of 0.4 unless stated otherwise.

    \subsection{1D relativistic shock tubes} 
    \label{sub:1d_relativistic_shock_tube}

      \begin{figure}
        \centering
        \includegraphics[width=0.5\textwidth]{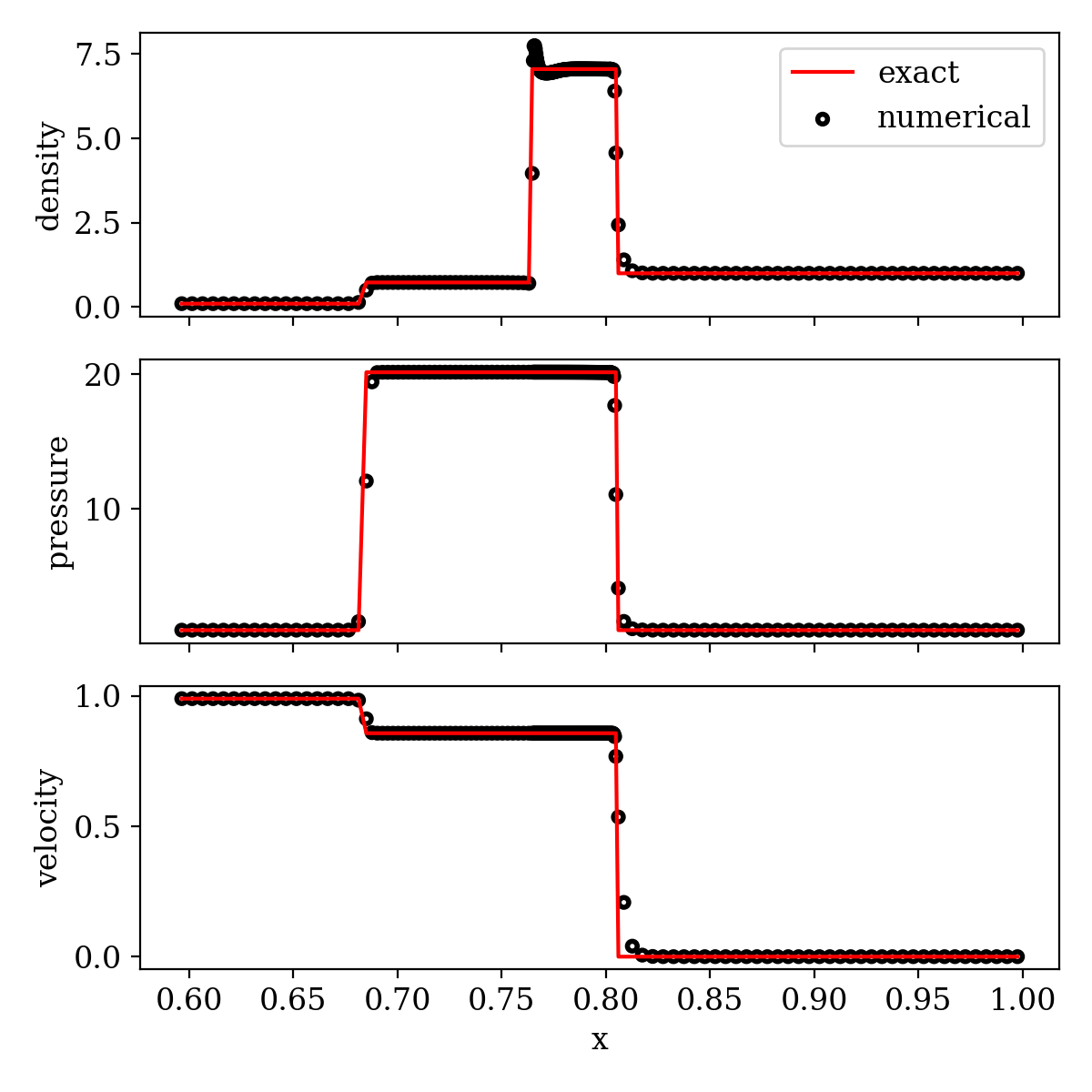}
        \caption{Shock tube in Cartesian coordinates. $t = 0.6$. Initial discontinuity at $x = 0.25$. Left state: $\rho = 0.1$, $p = 1$, $v = 0.99 \times c$. Right state: $\rho = 1$, $p = 1$, $v = 0$. Ideal EOS with $\gamma=4/3$. Initial resolution is 200 cells}
        \label{fig:1DST}
      \end{figure}

      \begin{figure}
        \centering
        \includegraphics[width=0.5\textwidth]{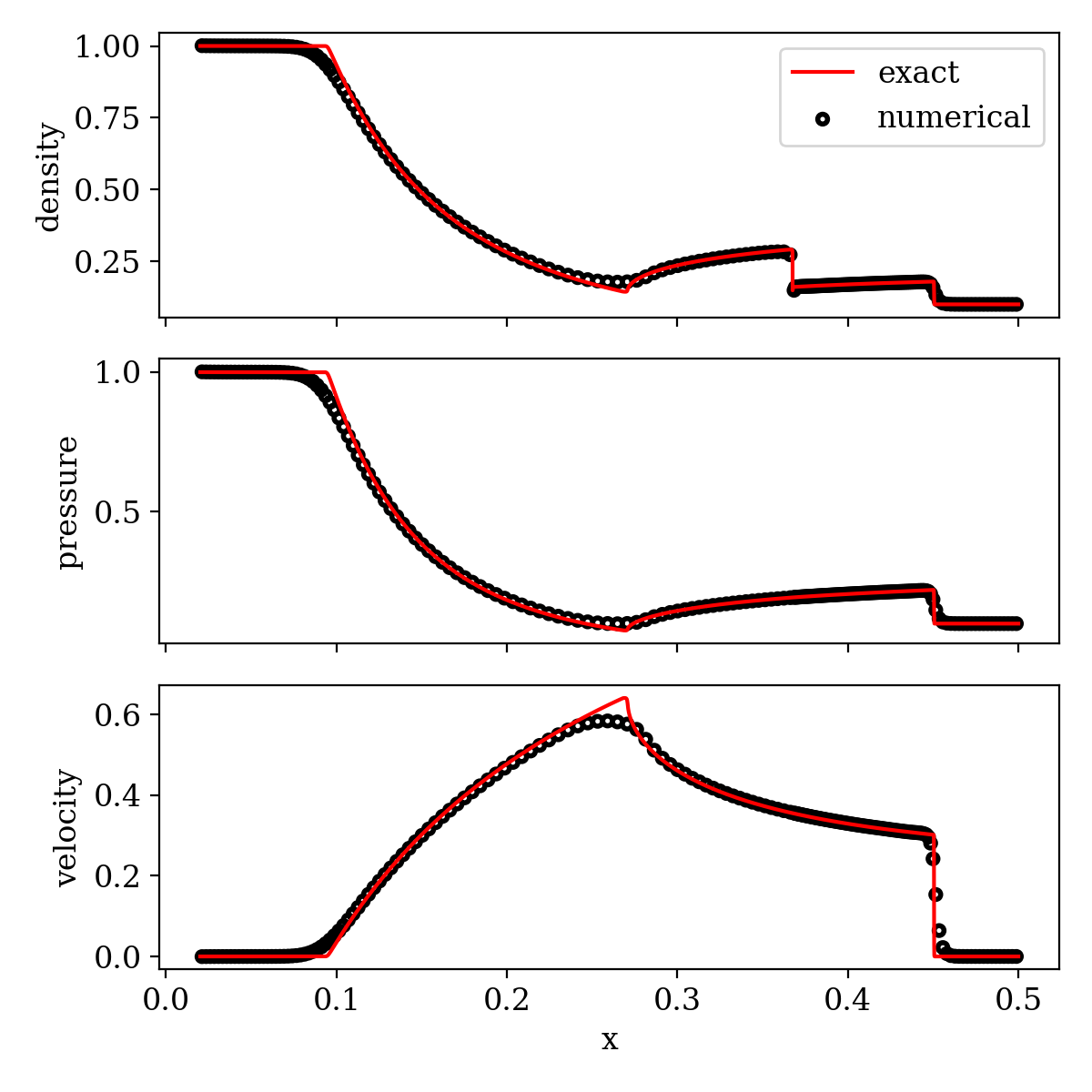}
        \caption{Shock tube in spherical coordinates. $t = 0.3$. Initial discontinuity at $x = 0.25$. Left state: $\rho = 1$, $p = 1$, $v = 0$. Right state: $\rho = 0.1$, $p = 0.1$, $v = 0$. Ideal EOS with $\gamma=4/3$. Initial resolution is 200 cells}
        \label{fig:1DST_sph}
      \end{figure}

      In figure \ref{fig:1DST} we show the result of a relativistic 1D shock tube in cartesian coordinates with the following parameters:
      \begin{align}
        (\rho, v, p) =
        \begin{cases}
          (0.1,0.99,1) & \text{for~} x \leq 0.25, \\
          (1,0,1) & \text{for~} x > 0.25,
        \end{cases}
      \end{align}
      where $v$ is in units of $c$. We also choose a fixed adiabatic index $\gamma=4/3$. 
      In figure \ref{fig:1DST_sph} we test our correct implementation of spherical coordinates and show the result of a relativistic 1D shock tube with parameters:
      \begin{align}
        (\rho, v, p) =
        \begin{cases}
          (1,0,1) & \text{for~} x \leq 0.25, \\
          (0.1,0,0.1) & \text{for~} x > 0.25.
        \end{cases}
      \end{align}

      In both systems of coordinates, the code is able to properly capture the shock positions as well as the contact discontinuity. The motion of the interfaces at the fluid velocity allows us to resolve contact discontinuity with only a few zones in these tests.


    \subsection{Isentropic wave} 
    \label{sub:isentropic_wave}

\begin{figure}
        \centering
        \includegraphics[width=0.5\textwidth]{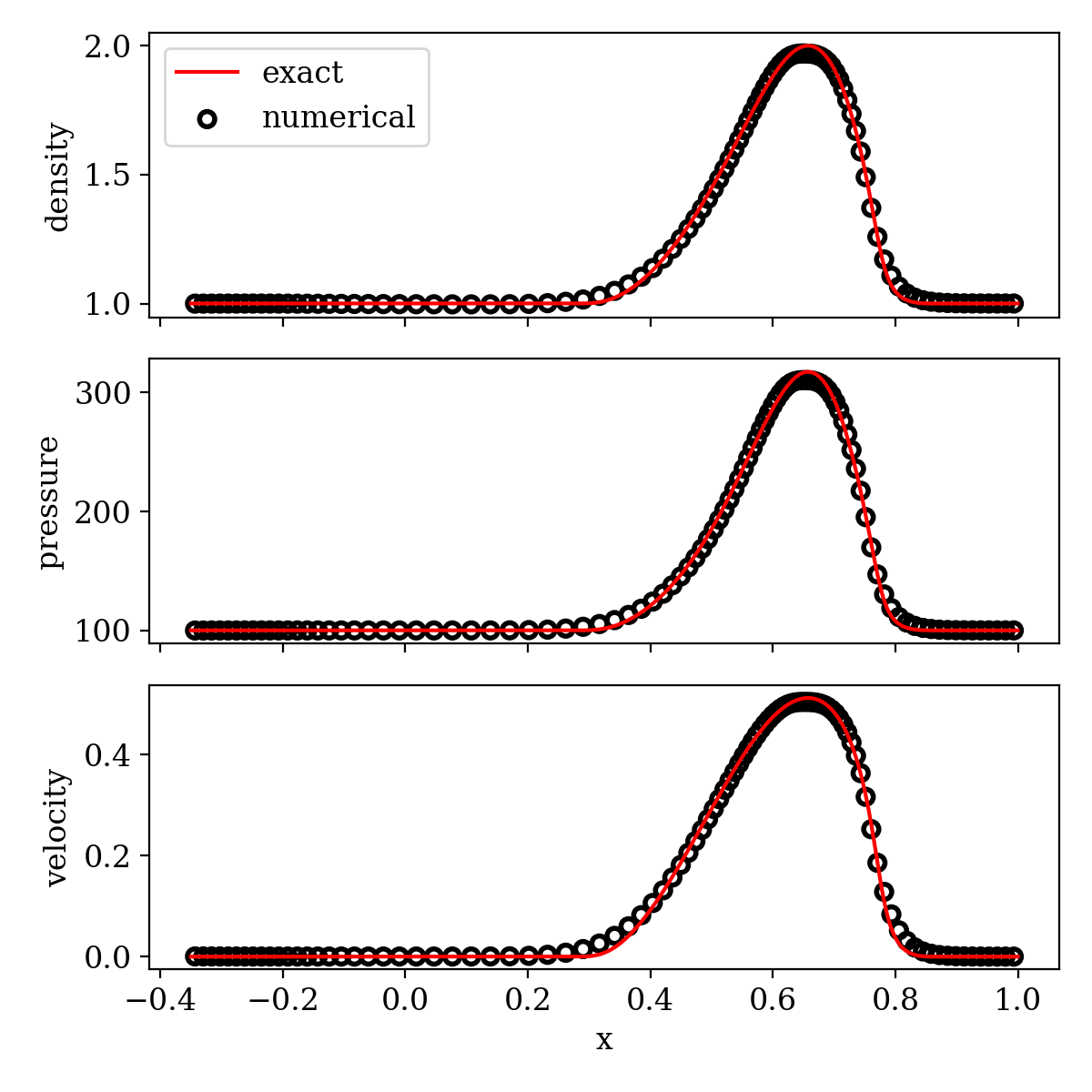}
        \caption{1D Isentropic wave test output for primitive variables with 100 fluid zones. The numerical results at t = 0.7 (black circles) are in very good agreement with the exact solution (red solid curve)}
        \label{fig:test_isen}
      \end{figure}

\begin{figure}
        \centering
        \includegraphics[width=0.5\textwidth]{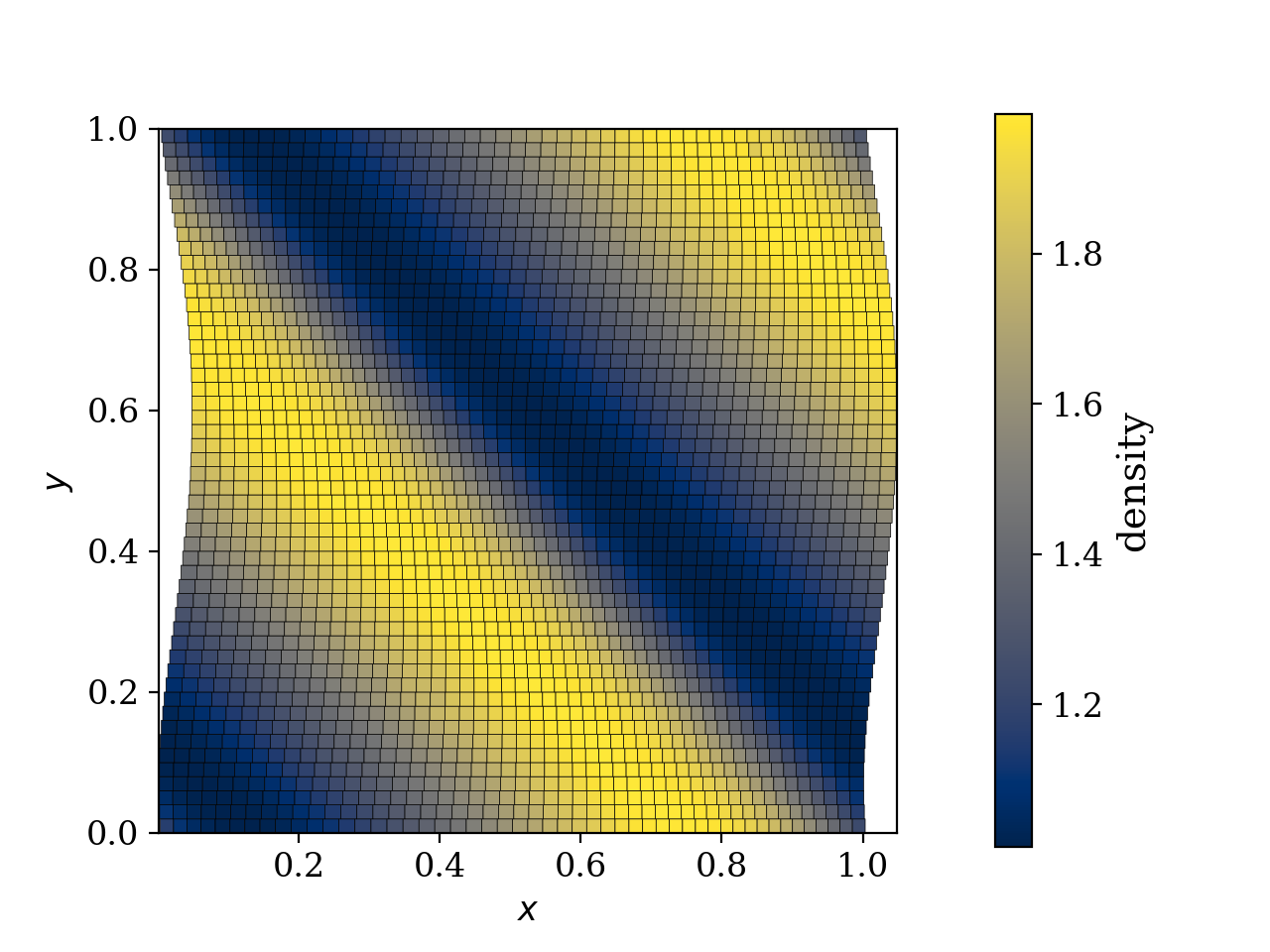}
        \caption{2D isentropic wave test at $t=0.7$ with initial uniform grid of resolution of $50\times50$. In this case, the mesh is allowed to move the x direction and distorts with the waves. The waves remain diagonal to the grid directions and follow the expected theoretical evolution, propagating towards increasing $x$ and $y$.}
        \label{fig:test_isen2D}
      \end{figure}

      We test the accuracy of our code for smooth regions of the flow by simulating the evolution of a 1D isentropic wave in cartesian coordinates. The setup we use is identical to that of \citet{Zhang2006}. We choose a fixed adiabatic index $\gamma=5/3$ for this setup. A comparison of the exact solution and the numerical result is shown in figure \ref{fig:test_isen}. The convergence rates at different resolutions are reported in table \ref{tab:conv_isentropic_wave}. 
      We nearly reach second order convergence for this test. We also assess the order of convergence of the code in two dimensions by running the isentropic wave in a direction diagonal to the initial grid in cartesian coordinates, as shown in figure \ref{fig:test_isen2D}. This setup is identical to the one from \citet{Duffell2011}. We choose periodic boundary conditions. We constrain the aspect ratio of the cells in the simulation domain to the $[0.5,2]$ interval. We allow the boundaries to move (this is compatible with the periodic boundary condition), which is responsible for the distortion of the grid visible in the figure. We report the convergence rates in table \ref{tab:conv_isen2D}.

      \begin{table}
        \centering
        \caption{Convergence analysis for the 1D isentropic wave test}
        \small
        \begin{tabular}{rcc}
          \hline
          \textbf{Resolution} & \textbf{L1 error} & \textbf{Convergence rate}\\
          \hline
            100  & 4.41e-3 &     \\
            316  & 5.87e-4 & 1.85\\
            1000 & 6.15e-5 & 1.92\\
            3160 & 6.92e-6 & 1.89\\
          \hline
        \end{tabular}
        \label{tab:conv_isentropic_wave}
      \end{table}

            \begin{table}
        \centering
        \caption{Convergence analysis for the 2D isentropic wave test}
        \small
        \begin{tabular}{rcc}
          \hline
          \textbf{Resolution} & \textbf{L1 error} & \textbf{Convergence rate}\\
          \hline
            20x20   & 1.94e-2 &     \\
            50x50   & 3.50e-3 & 1.89\\
            100x100 & 9.31e-4 & 1.91 \\
            300x300 & 1.16e-4 & 1.90\\
          \hline
        \end{tabular}
        \label{tab:conv_isen2D}
      \end{table}


    \subsection{2D Riemann problem} 
    \label{sub:2d_relativistic_shock_tube}
    
\begin{figure}
        \centering
        \includegraphics[width=0.5\textwidth]{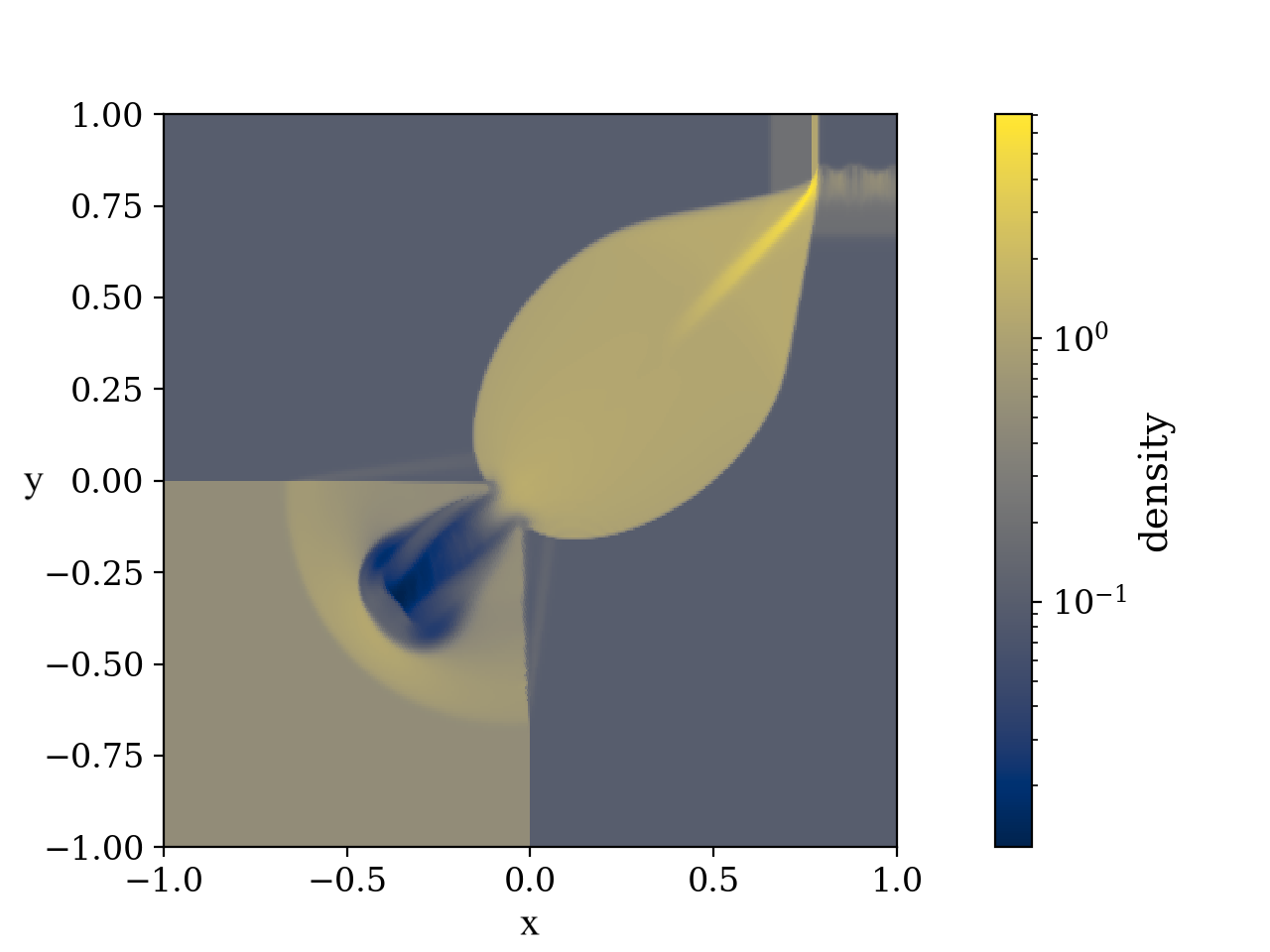}
        \caption{Density output for the 2D Riemann problem at $t=0.8$. Initial uniform grid of resolution $300\times300$. The mesh moves in the $x$ direction. Even though the mesh only moves in one direction, we observe a symmetrical evolution where the only difference between the $x$ and $y$ directions is the higher diffusion around the shocks in the non-moving direction.}
        \label{fig:2DST}
      \end{figure}

      To assess more complex 2D behavior of the code, we run a 2D Riemann problem with the same parameters as \citet{Mignone&Bodo2006}. This setup involves the interaction of four elementary waves formed at the interfaces between four initial different fluid states. On a square domain spanning $[-1,1]\times[-1,1]$, their setup is the following:
      \begin{align}
        (\rho, v_x, v_y, p) =
        \begin{cases}
          (0.1,0,0,0.01) & \text{for~} x,y > 0 \\
          (0.1,0.99,0,1) & \text{for~} x < 0 < y \\
          (0.5,0,0,1) & \text{for~} x,y < 0 \\
          (0.1,0,0.99,1) & \text{for~} y < 0 < x 
        \end{cases}.
      \end{align}
      In our setup, we use a resolution of $300\times300$ and allow the mesh to move in the $x$ direction and constrain the aspect ratio in the interval [0.1,1.5]. We show the output density at final time $t_f = 0.8$ in figure \ref{fig:2DST}. This test confirms the accuracy of the code in both the directions aligned and transverse to the mesh motion. It also highlights the increase in precision around shocks parallel to the mesh motion as features in the $x$ direction are more diffuse. The slight asymmetry in the region of lowest density is attributed to the difference in treatments between the x and y directions and nicely confirms the improvement in the direction of mesh motion. This region also particularly suffers from a high frequency of de-refinement operations on the lower-right edge, which is responsible for the loss in precision.

    \subsection{1D GRB jet - Blandford Mckee blast wave profile} 
    \label{sub:1d_simulation}

\begin{figure}
        \centering
        \includegraphics[width=0.5\textwidth]{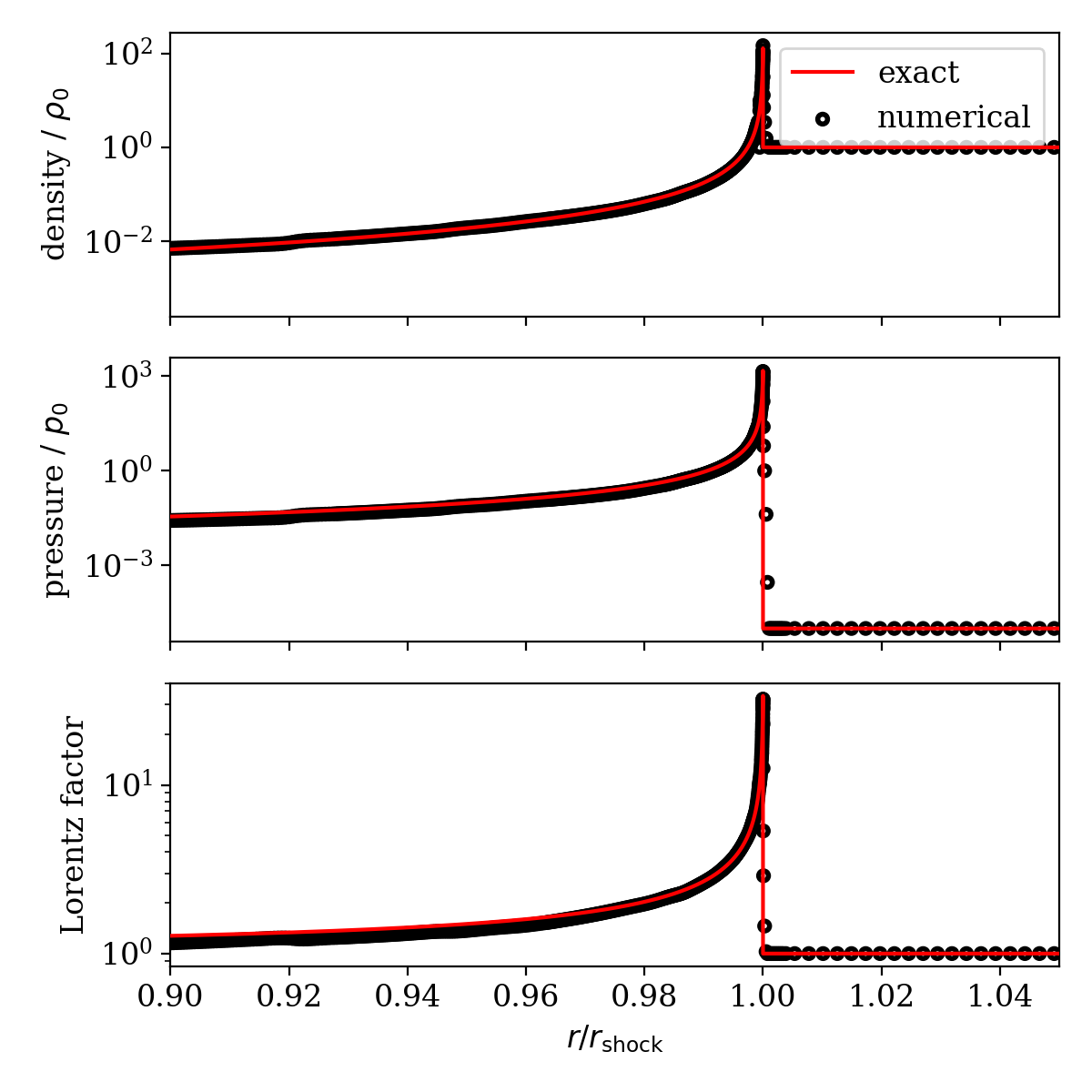}
        \caption{Blandford-McKee blast-wave output at $t=8.81\times10^{6}$s. $E_\mathrm{iso}=10^{53}$erg, initial peak fluid Lorentz factor $\Gamma=100$, CSM number density $n_0 = 1\mathrm{~cm}^{-3}$.}
        \label{fig:test_BM}
      \end{figure}

      A first application of this code to one-dimensional GRB blast wave simulations is done in \citet{Ayache2020}. It is important for any code applied to ultra-relativistic blast waves to demonstrate its ability to properly capture dynamics in these extreme regimes. As such an very important test is the comparison with the analytical solution for a relativistic point-like explosion, the Blandford-Mckee (BM) solution \citep{Blandford1976}. GRB blast waves transition to this asymptotic solution as they sweep up CSM material and it is surprisingly hard for fixed mesh AMR codes to properly capture the peak of the blast-wave, where particle acceleration happens.
      Here, we set up a BM solution at time $t_0$ and check that our numerical solution still matches the expected radial profile for the fluid quantities for $t>t_0$. We set up a blast wave with isotropic equivalent energy $E_\mathrm{iso}=10^{53}$erg and CSM uniform number density $n_0=1\mathrm{cm}^{-3}$ at an initial peak fluid Lorentz factor $\Gamma_0=100$ (initial time $t_0=4.36\times10^{6}s$. Figure \ref{fig:test_BM} shows the radial profile of primitive variables at $t=8.81\times10^{6}$s. The code accurately captures the shock position and the radial profile of the blast wave. Our code also fully captures the time evolution of the peak Lorentz factor at the shock front, which we demonstrate for two dimensions in section \ref{sub:synth_lc_results}. In this 1D test, we use the same AMR criteria as in the 2D simulations described section \ref{sub:synth_lc_params}, using a fiducial angular track width $\mathrm{d}\theta = \pi/2000$. We can see the advantage of the moving mesh approach where we can resolve the blast wave with little added computational cost as the region limiting the time-step is situated just ahead of the shock.

    \subsection{2D GRB jets - Rayleigh-Taylor instabilities in afterglows} 
    \label{sub:2d_simulation}

      Moving to two dimensions, we show the ability of the code to capture complex dynamics by investigating the growth of Rayleigh-Taylor (RT) instabilities at the contact discontinuity between the ejecta and the CSM. \citet{Duffell2013} \defcitealias{Duffell2013}{DM13}(hereinafter \citetalias{Duffell2013}) have already shown that RT instabilities can appear at the contact discontinuity in GRB afterglows by running moving mesh dynamical simulations. We reproduce here their approach and compare our results with those obtained using their code JET.

      \subsubsection*{Initial setup}
        We implement the fireball model \citep{Kobayashi1999} in which we input a given amount of energy $E_\mathrm{iso}$ and mass $M$ into a small sphere of radius $R_0$ placed in a uniform CSM of mass density $\rho_0$. At the initial time, the velocity of the fluid is 0 in the whole system. The thermal energy in the hot fireball is then converted to kinetic energy and the resulting blast wave will coast with fluid Lorentz factor $\Gamma = E_\mathrm{iso} / M$. Like \citetalias{Duffell2013} we place ourselves in the \emph{thin shell} limit where the initial structure of the fireball does not influence the later evolution. This is done by choosing $R_0$ small enough in order for the coasting and spreading phases to happen long before the deceleration phase: $\Gamma^2 R_0 \ll t_\gamma$, where $t_\gamma = (M/\Gamma \rho_0)^{1/3}$ is the deceleration time \citep{Kobayashi1999}. We run two simulations with $\Gamma \equiv 30$ (run30) and $\Gamma \equiv 100$ (run100). Setting $E_\mathrm{iso} \equiv 10^{52}$~erg determines the corresponding value of $M$. The rest of the initial parameters are reported in table \ref{tab:initial_parameters_RT}.

       \begin{table}
         \small
         \centering
         \caption{Initial parameters for the 2D GRB RT simulations}
         \begin{tabular}{llrl}
           \hline
           \textbf{Parameter} & \textbf{Notation} & \textbf{Value} & \textbf{Unit}\\
           \hline
           Equivalent Isotropic Energy  &  $E_\mathrm{iso}$  & $10^{52}$ & erg \\
           Coasting Lorentz factor  &  $\Gamma$ & $30$;$100$     &       \\
           Initial radius of the fireball & $R_0$ & 100;0.4 & l.s.\\
           CSM number density  & $n_0$   & $1$   & $\mathrm{cm}^{-3}$\\
           Temperature of CSM ($p/\rho c^2)$ &  $\eta$ & $10^{-5}$ & \\
           \hline
         \end{tabular}
         \label{tab:initial_parameters_RT}
       \end{table}

      \subsubsection*{1D early run and grid parameters}
        All simulations are carried out in axisymmetric spherical coordinates $(r, \theta, \phi)$. For the sake of computational efficiency, we first run 1D simulations of the acceleration and coasting phases of the fireball, before deceleration. 1D simulations are sufficient in this regime since the collimated jet is not yet causally connected and we can thus assume spherical symmetry. Transverse motion will appear with the instabilities after the deceleration time and we will need to transition to 2D before then. We initialise the fireball on a logarithmic radial grid with 600 cells. The inner boundary is initially placed at $0.01~R_0$ and set to reflective boundary conditions. After the acceleration phase, we set the inner boundary velocity to $0.5~c$ and outflow boundary conditions to reduce the computation grid size. The outer boundary moves at $1.05~c$ throughout the whole simulation. We call $r_\mathrm{max}$ this increasing outer radius. In the 2D stage, we set reflective boundary conditions at $\theta = 0$ and $\theta = \theta_\mathrm{simu}$. 

        We transition to the 2D simulations at $t\sim0.5~t_\gamma$, by broadcasting the result of the 1D solution onto $N_\theta \equiv 200$ radial tracks evenly distributed in the interval $\theta \in [0, \theta_\mathrm{simu} \equiv \pi/32]$. Since our simulation domain is half as wide, this corresponds to the 400 tracks case in \citetalias{Duffell2013}. We run the simulation until $t_M = (M/\rho_0)^{1/3}$ which is the time at which the blast-wave becomes non-relativistic. In both the 1D and 2D stages, the radial resolution is governed for each cell by a modified cell aspect ratio criterion $\hat{a} = \mathrm{d}r / (r_\mathrm{max}\mathrm{d}\theta)$. We choose to use $\hat{a}$ instead of the actual aspect ratio $a = \mathrm{d}r / (r\mathrm{d}\theta)$ to prevent the time-step from being limited by the cells located at small radii at later stages of the evolution. For each cell, $\hat{a}$ is allowed to vary in the interval $\hat{a} \in [0.2,5]$. We normalise the pressure and density by $\rho_0$ and $\eta \rho_0 c^2$, respectively. Throughout the evolution we floor the normalised density and pressure to $10^{-10}$.

      \subsubsection*{Results}

        \begin{figure*}
          \begin{multicols}{2}
            \includegraphics[width=0.45\textwidth]{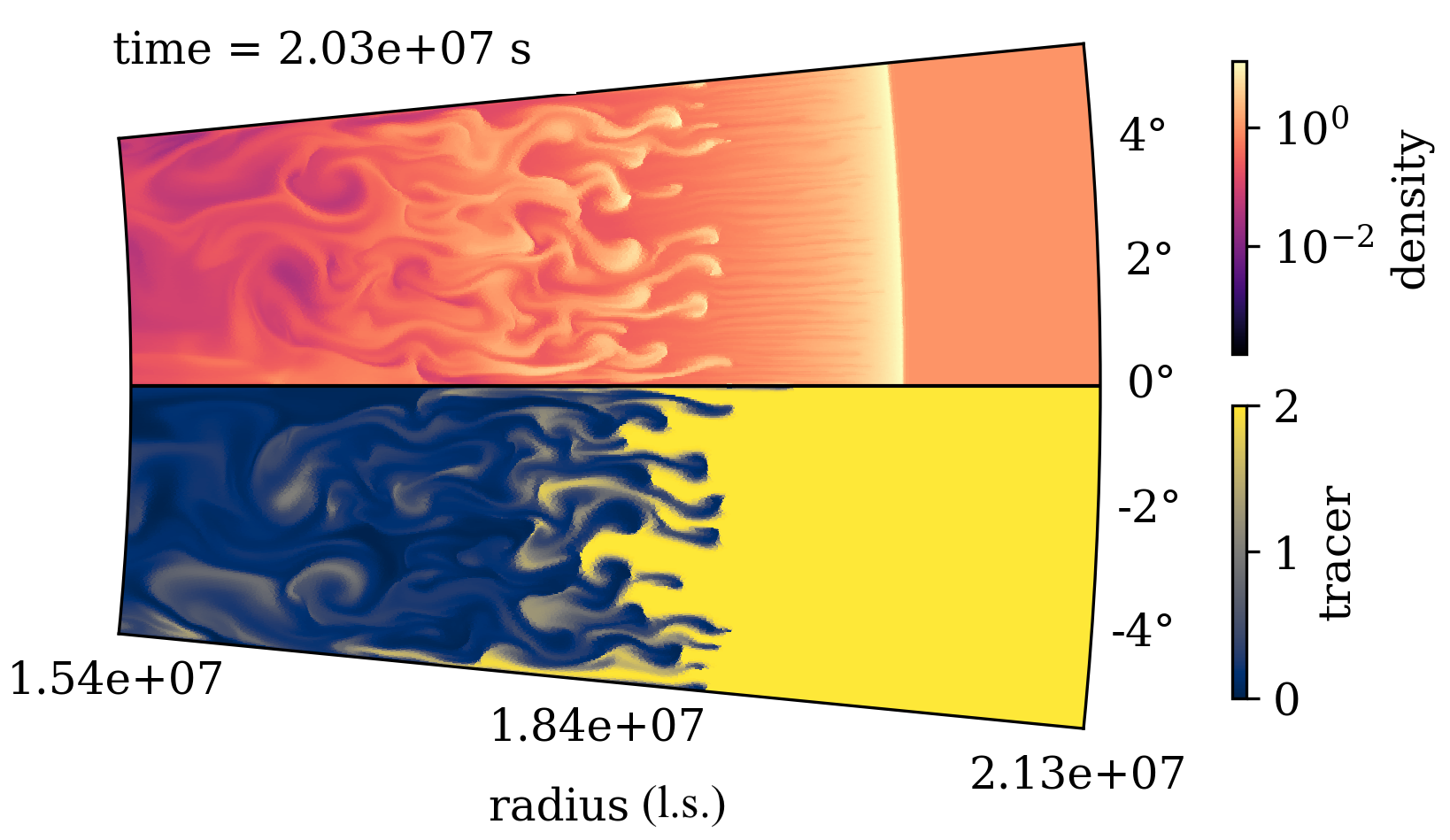}\par
            \includegraphics[width=0.45\textwidth]{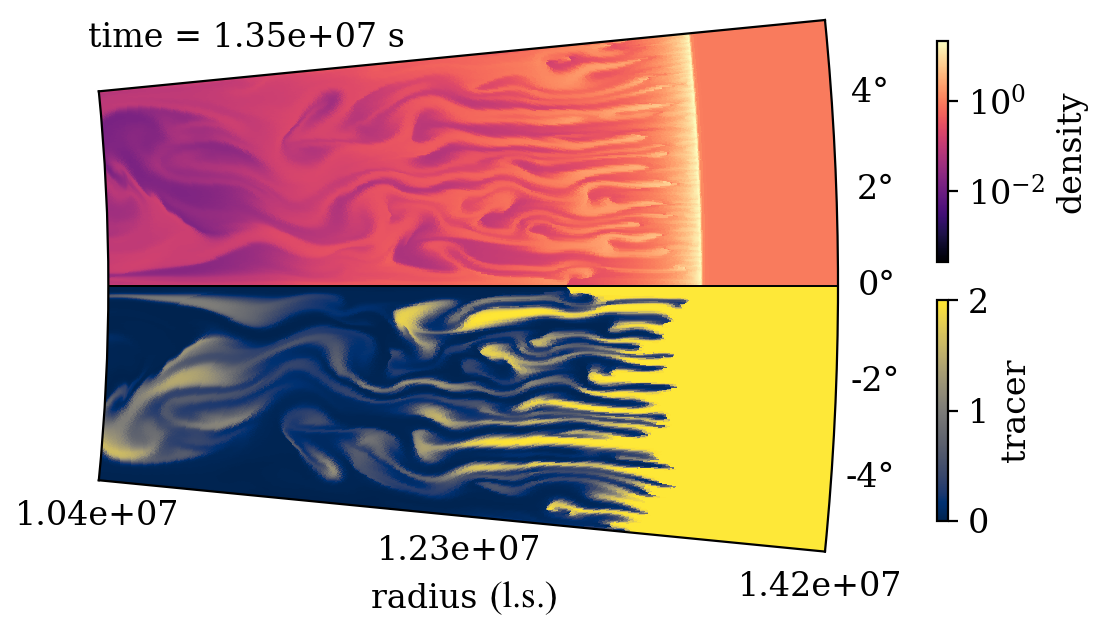}\par
          \end{multicols}
          \caption{2D Snapshots at $t=t_M$ of run30 (left) and run100 (right). The density is normalised by $\rho_0$. The tracer highlights the mixing between ejecta material (tracer=0) and CSM material (tracer=2).}
          \label{fig:RT_snapshots}
        \end{figure*}

        \begin{figure*}
          \centering
          \begin{multicols}{2}
            \includegraphics[width=0.45\textwidth]{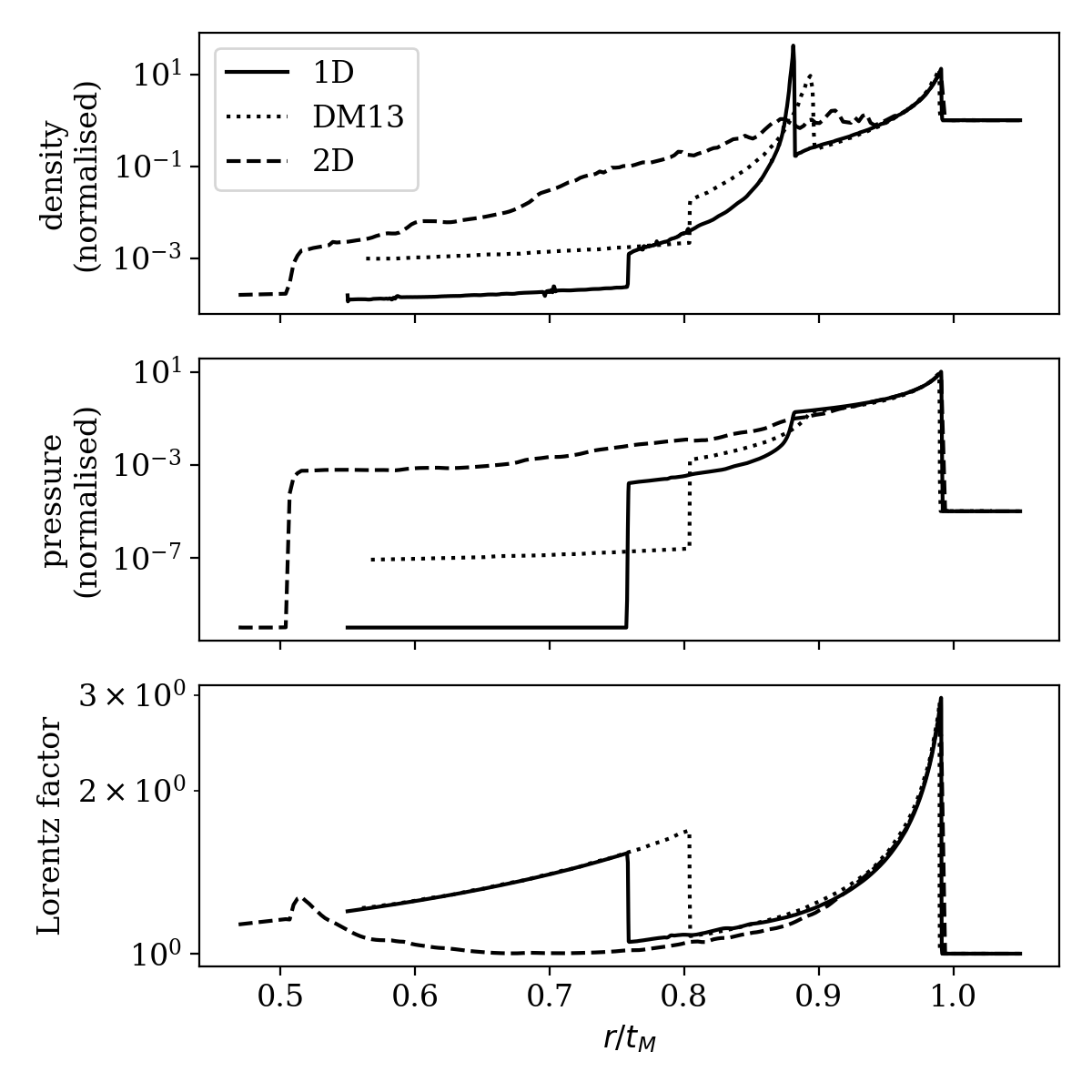}\par
            \includegraphics[width=0.45\textwidth]{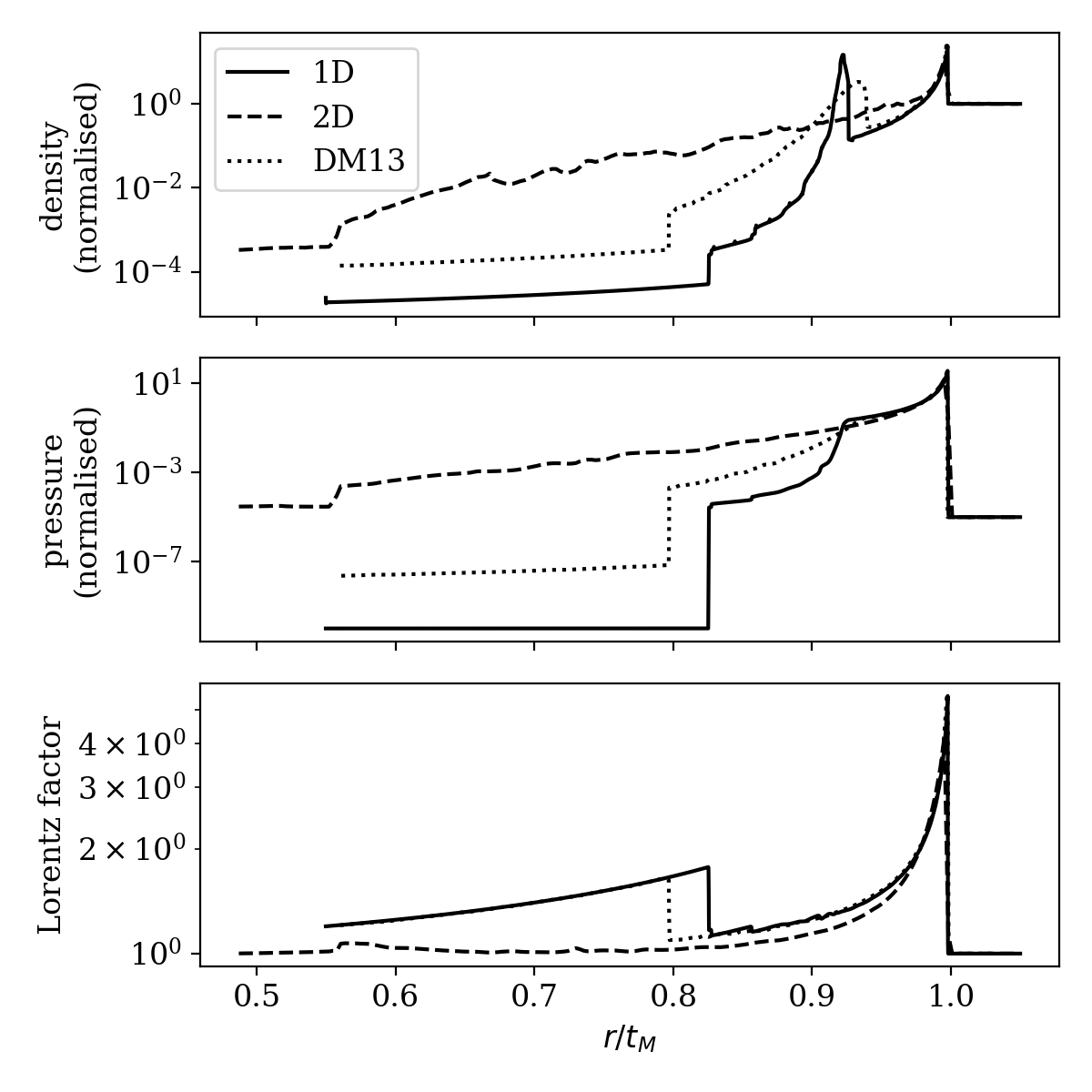}\par
          \end{multicols}
          \caption{Radial profiles. Density and pressure are normalised by the external medium density value. The DM13 curve is their 1D solution. In all cases, the forward shock is very accurately captured. In 2D, the RT instabilities push the reverse shock further at the back of the ejecta.}
          \label{fig:RT_radial_profiles}
        \end{figure*}

        Snapshots at $t_M$ for run30 and run100 are shown in figure \ref{fig:RT_snapshots} and the corresponding radial profiles are reported in figure \ref{fig:RT_radial_profiles}. We obtain very similar results to \citetalias{Duffell2013}. The radial profiles in 2D were obtained by averaging the fluid quantities over the $\theta$ direction weighing track contributions by their respective volumes. We also ran 1D simulations up to $t_M$ to compare with the results from the 2D runs. First looking at these radial profiles, our forward shock (FS) closely coincides with the FS in the simulations of \citetalias{Duffell2013}. However, we notice a discrepancy in the reverse shock (RS) position and a difference in the rest-mass density behind the ejected material. This not linked to our different choice of EOS (they use an ultra-relativistic ideal gas with fixed $\gamma=4/3$, as opposed to our trans-relativistic EOS) as the difference is still visible when switching to their fluid description. We have confirmed that mass was conserved in the ejecta using the passive scalar tracer we have set up. Finally we ran the code for various values of $R_0$ and noticed no significant difference between our runs. \citetalias{Duffell2013} do not specify the treatment of their inner boundary.

        The 2D runs exhibit the same features as those found in \citetalias{Duffell2013}. We observe Rayleigh-Taylor instability at the contact discontinuity growing from the numerical noise without the need for seeding. We can also confirm the influence of this instability on the position of the RS. As the instability grows, the turbulence is able to reach the RS and pushes it faster towards the back of the ejecta. This effect is particularly well visible in run30 in figure \ref{fig:RT_radial_profiles} and so strong in run100 that the RS has actually already left the simulation domain at $t_M$. In order to properly quantify this effect, we would need to run higher resolution simulations like in \citetalias{Duffell2013} as the transverse size of the tracks currently prevents the instability from growing properly. While sufficient resolution is almost achieved in run30, we do not observe the smaller scales expected at higher Lorentz factors. In run100 the maximum angular size at which RT instability can develop is 1/150 rad, corresponding to 13.6 angular track widths. It is therefore likely that smaller scale instability was not sufficiently resolved.

        These results confirm the ability of the code to properly capture complex relativistic dynamics on a moving mesh. They also confirm that the CD can be unstable in the afterglow. Since this particular phenomenon has previously been explored in the literature, we refer the reader to \citetalias{Duffell2013} for more in depth analysis.


  \section{Local synchrotron cooling} 
  \label{sub:local_synchrotron_cooling}

    We have shown in the previous two sections that our code is able to properly capture complex relativistic hydrodynamics on a moving mesh. In this section, we describe our approach to implementing local tracing and radiation of particles accelerated at the shock fronts in the fluid.

    \subsection{Tracing of accelerated particles} 
    \label{sub:tracing_of_accelerated_particles}
    

    We follow the prescription first described in \citet{Downes2002} and \citet{VanEerten2010} and implemented on a moving mesh in one dimension in \citet{Ayache2020}. We expand on this previous implementation by including a description of the spectral evolution of the particle population that depends on shock velocity.

    Shock waves are responsible for the acceleration of electrons that radiate in synchrotron across the whole electromagnetic spectrum. We model this accelerated population with a truncated power-law in energy \citep{Sari1997a,Wijers1999}:
    \begin{align}
      n_e^{\prime}(\gamma_e^{\prime}) \propto 
      \begin{cases}
        (\gamma_e^{\prime})^{-p} & \text{ if } \gamma_\mathrm{min} < \gamma_e^{\prime} < \gamma_\mathrm{max} ,\\
        0 & \text{ otherwise},
      \end{cases}
    \end{align}
    where primed quantities are expressed in the co-moving frame. $n_e^\prime$ is the spectral number density as a function of $\gamma_e^\prime$ the Lorentz factor of an electron. $\gamma_\mathrm{min}$ is the minimum Lorentz factor (of the bulk of the population) and $\gamma_\mathrm{max}$ is the maximum Lorentz factor that decreases as the population cools. $p$ is the spectral index (power-law slope of this population). The distribution is normalised by considering that a fraction $\xi_N$ of electrons are accelerated and carry a fraction $\epsilon_e$ of the internal energy density $e = \rho \epsilon$. The cooling of these electrons is driven by synchrotron losses and adiabatic expansion:
    \begin{align} 
      \label{eq:cooling}
      \frac{\mathrm{d}\gamma_e^\prime}{\mathrm{d}t^\prime} &=
        - \frac{\sigma_T (B^\prime)^2}{6 \pi m_e c } (\gamma_e^\prime)^2
        + \frac{\gamma_e^\prime}{3 \rho}\frac{\mathrm{d}\rho}{\mathrm{d}t^\prime},
    \end{align}
    with $B^\prime = \sqrt{8 \pi \epsilon_B e}$ the local magnetic field intensity derived from magnetic energy expressed as a fraction $\epsilon_B$ of internal energy. $m_e$ and $\sigma_T$ are the electron mass and the Thompson cross section, respectively. This expression can be re-cast into an advection equation:
    \begin{align}
    \label{eq:advect_cool}
    \frac{\partial}{\partial t} \left( \frac{\Gamma \rho^{4/3}}{\gamma_e^\prime} \right) + \frac{\partial}{\partial x^{i}} \left( \frac{\Gamma \rho^{4/3}}{\gamma_e^\prime} v \right)
      = \frac{\sigma_T}{6 \pi m_e c} \rho^{4/3} (B^\prime)^{2},
  \end{align}
  that the hydrodynamics solver can treat as a passive scalar with a source term. The bounds of the population of accelerated electrons are locally evolved downstream of shocks following this procedure. Particle injection at shock fronts is simply done by resetting the values of $\gamma_\mathrm{max}$ and $\gamma_\mathrm{min}$. $\gamma_\mathrm{max}$ directly downstream of the shock is theoretically set by the acceleration time-scale. For sufficiently large $p$ it can be taken to be infinity. In practice, we just set $\gamma_\mathrm{max}=10^8$ to a high enough value such that the frequency cut-off in the observer frame is higher than $10^{18}$~Hz, the highest frequency at which we compute the radiation. $\gamma_\mathrm{min}$ is set by normalising the total available energy over the electron population:
  \begin{align}
    \gamma_\mathrm{min} = \frac{p-2}{p-1} \frac{\epsilon_e e}{n' m_e c^2},
  \end{align}
  where $n^\prime$ is the accelerated electron number density in the co-moving frame.

  We have also added in the radiative code the possibility of using a local value of $p$. Indeed, most works currently assume fixed spectral index $p \sim 2-2.5$. However, we expect shock strength to vary during the dynamical evolution, which leads to a varying spectral index in the accelerated population. To evaluate the effect of this evolution we implement this in our shock detector and subsequently advect the spectral index value $p$ as a simple passive scalar field:
  \begin{align}
    \frac{\partial}{\partial t} (\Gamma \rho p) + \frac{\partial}{\partial x^{i}} (\Gamma \rho p v) = 0.
  \end{align}
  The initial value of p is chosen following \citet{Kirk2000} and \citet{Keshet2005} (see e.g. \citet{Sironi2015} and \citet{Marcowith2020} for recent reviews on particle acceleration in relativistic shocks), where we have identified the upstream fluid velocity with the bulk fluid velocity. The dependency of $p$ on the upstream velocity comes from the induced anisotropy upstream of the shock for the electron population for a relativistic shock (the so-called "spectrum - anisotropy connection").
  This approximation greatly simplifies the implementation and is acceptable as we will only be interested in forward shock emission from GRB afterglows in this work, but a different approach would be needed for e.g. reverse shock contribution or internal shocks.
  The dependency of $p$ on the upstream fluid four-velocity $u$ is approximated by the following expression in our code:
  \begin{align}
    p = 2.11 + 0.11 \times \tanh(\log_{10}(u/3.16)).
  \end{align}

  \subsection{Deep Newtonian phase} 
  \label{sub:deep_newtonian_phase}

    The approach to initializing and evolving $\gamma_\mathrm{min}$ as described in the previous section will at very late times inevitably lead to a non-physical value $\gamma_\mathrm{min}$ (the issue is exacerbated if $p$ is allowed to approach 2): the total available energy can no longer be stored in the non-thermal electron population while keeping $\gamma_\mathrm{min}$. Furthermore, if $\gamma_\mathrm{min}$ gets too close to 1, our description of the particle population as a power-law in energy breaks down. 

    \citet{VanEerten2010} and \citet{Sironi2013} \defcitealias{Sironi2013}{SG13}(hereinafter \citetalias{Sironi2013}) address this problem by varying the fraction of accelerated electron. In theory, the electron population should be described in the Newtonian phase by a power-law in momentum \citep{Bell1978,Blandford1978,Blandford1987}. Lowering the electron participation fraction mimics this behaviour by keeping the energy in the non-thermal population contained above $\gamma_\mathrm{min} \beta_\mathrm{min}\sim 1$ (where $\beta_\mathrm{min}$ is the electron velocity in terms of c) through shifting the lower cut-off Lorentz factor upwards to higher values. In this work, we adapt the prescription from \citetalias{Sironi2013} to our local description of the particle population. This approach has the advantage of being compatible with our prescription for the local radiative cooling presented above, while still manifesting itself in the light curve at the correct time.

    The accelerated electron population can be split into a relativistic population that contributes to radiation and a non-relativistic population for which the energy and the radiative contribution are considered negligible. We write $\xi_{N,\mathrm{DN}}$ the fraction of relativistic electrons. Because we account for the impact of synchrotron energy losses when locally tracing the evolution of $\gamma_\mathrm{min}$, we cannot switch over to the adiabatic prescription from \citetalias{Sironi2013} wholesale. Instead, we recover the total energy in the electron population from the integrated energy between $\gamma_\mathrm{min}$ and $\gamma_\mathrm{max}$. For this purpose, we still mathematically allow $\gamma_\mathrm{min}$ to drop below unity when locally tracing its value (but cap it at 1 when computing the corresponding critical frequency). The population between $\gamma_\mathrm{min}$ and 1 is considered cold and ceases to contribute to the emission. We have:
    \begin{align}
      E'_\mathrm{tot} &= \int_{\gamma_\mathrm{min}}^{\gamma_\mathrm{max}} n_e'(\gamma_e') \gamma_e' m_e c^2 \mathrm{d}\gamma_e' \\
      &= \int_{\gamma_\mathrm{min}}^{\gamma_\mathrm{max}} C_\mathrm{tot} (\gamma_e')^{1-p} m_e c^2 \mathrm{d}\gamma_e',
    \end{align}
    where $C_\mathrm{tot}$ is the population normalisation factor if the energy were stored between $\gamma_\mathrm{min}$ and $\gamma_\mathrm{max}$.
    Equating the local energy $E'_\mathrm{tot}$ computed from the numerical simulations and the energy in the relativistic part yields a first equation:
    \begin{align}
      \int_{\gamma_\mathrm{min}}^{\gamma_\mathrm{max}} C_\mathrm{tot} (\gamma_e')^{1-p} \mathrm{d}\gamma_e'
       = \int_{1}^{\gamma_\mathrm{max}} C_\mathrm{rel} (\gamma_e')^{1-p} \mathrm{d}\gamma_e',
       \label{eq:energy_DNP}
    \end{align}
    where $C_\mathrm{rel}$ is the relativistic population normalisation factor for which the energy $E'_\mathrm{tot}$ is stored between $1$ and $\gamma_\mathrm{max}$. By using the normalisations for the electron number density:
    \begin{align}
      n' = \int_{\gamma_\mathrm{min}}^{\gamma_\mathrm{max}} C_\mathrm{tot} (\gamma'_e)^{-p} \mathrm{d}\gamma'_e,
    \end{align}
    and
    \begin{align}
      \xi_{N,\mathrm{DN}} \times n' = \int_{1}^{\gamma_\mathrm{max}} C_\mathrm{rel}(\gamma'_e)^{-p}\mathrm{d}\gamma'_e,
    \end{align}
    where $n'$ is the accelerated electron number density, one can eliminate $C_\mathrm{tot}$ and $C_\mathrm{rel}$ from eq. \ref{eq:energy_DNP}, and obtain the relativistic electron fraction:
    \begin{align}
      \xi_{N,\mathrm{DN}} = \frac{\gamma_\mathrm{max}^{2-p} - \gamma_\mathrm{min}^{2-p}}{\gamma_\mathrm{max}^{2-p}-1} \times \frac{\gamma_\mathrm{max}^{1-p}-1}{\gamma_\mathrm{max}^{1-p}- \gamma_\mathrm{min}^{1-p}}.
      \label{eq:DNP}
    \end{align}

    During our calculation of the emissivity, in the post-processing of the hydrodynamical snapshots, whenever we encounter $\gamma_\mathrm{min} < 1$, we compute the corresponding value of $\xi_{N,\mathrm{DN}}$ and then set $\gamma_\mathrm{min} \equiv 1$. We then compute the emissivity from the relativistic electrons by using $\xi_{N,\mathrm{DN}}n'$ instead of $n'$ in the steps decribed in the following section.


  \subsection{Radiative flux calculation} 
  \label{sub:radiative_flux_calculation}
  
    To calculate the corresponding flux, we have adapted the linear radiative transfer approach and corresponding code from \citet{VanEerten2009a} using the same simplified connected power-law description for synchrotron emission as in \citet{VanEerten2010a}. We intend to upgrade the code in the future with more a elaborate description including a treatment of the transition between these power-law regimes as described in \citet{VanEerten2009a}. For the sake of simplicity we neglect self-absorption here and focus on frequencies above the radio band.

    The frequency at which a single electron with energy $\gamma_e^\prime m_e c^2$ in the co-moving frame produces synchrotron radiation is $\nu_\mathrm{syn}(\gamma_e^\prime) = \frac{3q_e B^\prime}{16 m_e c} (\gamma_e^\prime)^2$ with $q_e$ the charge of the electron. We can now write $\nu_\mathrm{min}^\prime = \nu_\mathrm{syn}(\gamma_\mathrm{min})$, and $\nu_\mathrm{max}^\prime = \nu_\mathrm{syn}(\gamma_\mathrm{max})$. The spectral volumetric power for an emitting region of the fluid is given by:
    \begin{align}
      P_\nu^\prime = 
      \begin{cases}
        P_{\nu,\mathrm{max}}^\prime \left( \frac{\nu}{\nu_\mathrm{min}^\prime} \right)^{1/3}, & \nu < \nu_\mathrm{min}^\prime, \\
        P_{\nu,\mathrm{max}}^\prime \left( \frac{\nu}{\nu_\mathrm{min}^\prime} \right)^{-(p-1)/2}, & \nu > \nu_\mathrm{min}^\prime, 
      \end{cases}
    \end{align}
    \begin{align}
      &\text{with~} P_{\nu,\mathrm{max}}^\prime = \frac{4(p-1)}{3p - 1} 
        \times n^\prime \sigma_T \frac{4}{3} \frac{B^\prime}{6\pi} \frac{16m_e c}{3q_e}.
    \end{align}

    To account for local cooling, we implement a simple sharp cut-off in the spectral emissivity for frequencies above $\nu_\mathrm{max}^\prime$ in the fluid frame. In theory the emitted radiation follows and exponential cutoff that will be implemented in the future. We expect from our prescription a small underestimation of the true flux above the cooling break. This does not change the interpretation of the results from section \ref{sub:synthetic_grb_afterglow_light_curves_with_local_cooling_from_2d_simulations} as we will see the flux is generally higher with our approach compared to previous prescriptions. Eventually, once having taken into account the proper beaming factors depending on $\mu$ the cosine of the angle between the fluid velocity and the observer, and accounting for photon arrival times, the flux received at a given observer time $t_\mathrm{obs}$ for a given frequency $\nu$ can be integrated following:
    \begin{align}
      F(\nu, t_\mathrm{obs}) = \frac{1+z}{2d_L^2}\int_{-1}^1\mathrm{d}\mu\int_0^\infty r^2 \mathrm{d}r \frac{P^\prime_{\nu^\prime}(r, t_\mathrm{obs}+r\mu)}{\Gamma^2 (1 - \beta\mu)^2},
    \end{align}
    with $d_L$ the luminosity distance and $z$ the redshift.



  \subsection{Shock detection algorithm} 
  \label{sub:shock_detector}

    This radiative prescription relies on accurate detection of the shock positions in the fluid. While \citet{Ayache2020} make use of a shock detector based on the calculation of the limiting relative velocities at cell interfaces in 1D from \citet{Rezzolla2003a} and \citet{Zanotti2010}, we use in this work the more complex multi-dimensional version of this dectector, introduced by the same authors, that we describe in this section.

    Let us consider a candidate discontinuity in fluid quantities in which we observe a jump in density, pressure and velocity between two regions denoted 1 and 2. This is the setup of a local Rieman problem for which we can compute a criterion on the relative velocity orthogonal to the discontinuity $v_{12} \equiv (v_1 - v_2) / (1-v_1 v_2)$ for the formation of one shock and one rarefaction ($\mathcal{SR}$), or two shocks ($2\mathcal{S}$) in the resulting Riemann fan. By checking for this criterion on all the interfaces of the grid we can map the location of the shocks in all directions.

    In the $\mathcal{SR}$ case, the criterion is given by:
    \begin{align}
      v_{12} > (\tilde{v}_{12})_\mathcal{SR} 
        = \tanh \left( \int_{p_1}^{p_2} 
          \frac{\sqrt{h^2 + \mathcal{A}_1^2 (1-c_s^2)}}{(h^2+\mathcal{A}_1^2) \rho c_s}
          \mathrm{d}p
          \right),
    \end{align}
    where $c_s$ is the speed of sound and $\mathcal{A}_1 \equiv h_1 \gamma_1 v_1^t$, with $v_1^t$ the transverse velocity. We compute $(\tilde{v}_{12})_\mathcal{SR}$ by numerical integration.

    In this simple form, the algorithm does not discriminate on shock strength and can lead to the spurious detection of weak shocks. This can be adjusted by computing the limiting relative velocity for the $2\mathcal{S}$ case:
    \begin{align}
      v_{12} > (\tilde{v}_{12})_{2\mathcal{S}}
        = \frac{(p_1 - p_2) (1 - v_2 \bar{V}_s)}
        {(\bar{V}_s - v_2) \{ h_2 \rho_2 \gamma_2^2 [1 - v_2^2] + p_1 - p_2 \}}.
    \end{align}
    We refer the reader to \citep{Rezzolla2003a} for the explicit expression of $\bar{V}_s$ which is the velocity of $\mathcal{S}_\rightarrow$ in the Riemann fan $\{1\mathcal{S}_\leftarrow 3 \mathcal{C} 3^\prime \mathcal{S}_\rightarrow 2\}$ in the limit case where $p_3 \rightarrow p_1$.

    The shock detection threshold is adjusted by computing a new limit $(\tilde{v}_{12})_\mathrm{eff}$ with adjustable parameter $\chi \in [0,1]$ such that a shock is detected for:
    \begin{align}
      v_{12} > (\tilde{v}_{12})_\mathrm{eff} = (\tilde{v}_{12})_\mathcal{SR} 
        + \chi [(\tilde{v}_{12})_{2\mathcal{S}} - (\tilde{v}_{12})_\mathcal{SR}].
    \end{align}
    We find $\chi = 0.5$ produces satisfactory results in all the applications presented in this paper.

    This procedure can be generalized to any spacetime metric (or system of coordinates) by projecting the velocities into a local tetrad following $v^{\hat{i}} = M_j^{\hat{i}}v^j$ \citep{Pons1998}. In this work we are only interested in spherical coordinates for which the projection simply reduces to $M_j^{\hat{i}} = \mathrm{diag}(1, r, r\sin\theta)$.


\section{Synthetic GRB afterglow light curves with local cooling} 
\label{sub:synthetic_grb_afterglow_light_curves_with_local_cooling_from_2d_simulations}

  \subsection{Initial setup}

  We run simulations of spreading top-hat jets with local and global cooling. The setups are similar to those from \citet{VanEerten2012}. We start from a BM solution constrained to a small opening angle. This ensures that the results can be re-scaled making use of scale invariance with regard to the ratio of burst energy over circumburst medium density. The initial time $t_0$ is chosen such that the peak fluid Lorentz factor of the outflow is set to $\Gamma_\mathrm{peak}=100$. We run two simulations with the same dynamical parameters but different micro-physical parameters leading to a slow-cooling and fast-cooling early times case for us to analyse. Since we locally compute the microphysics these parameters indeed need to be specified before the dynamical simulation. Of course, the fast-cooling case is expected to transition to slow-cooling at later times and these different sets of micro-physical parameters are only selected to make our interpretation more straightforward. The micro-physical parameters also sit within the distributions derived from GRB observations \citep[see e.g.][]{Santana2014,Beniamini2015,Aksulu2021}. In this regard, the choice of $\epsilon_B \equiv 0.1$ in the fast-cooling case sits in the upper bracket of the distribution. The qualitative effect of the local cooling will however remain the same for all fast-cooling spectra regarless of the value of $\epsilon_B$. All the initial parameters in these simulations are reported in table \ref{tab:initial_parameters}.
  The simulation final time $t_f$ is determined by the time spanned by the synthetic light-curve. In practice, we choose $t_f = 3.33 \times 10^8$s and we check that the last snapshot does indeed not contribute to the emission at the final observer time $t_\mathrm{obs,f} = 10^8$s. 

  \begin{table}
    \small
    \centering
    \caption{Initial parameters for the 2D GRB jet local cooling simulation}
    \begin{tabular}{llrl}
      \hline
      \textbf{Parameter} & \textbf{Notation} & \textbf{Value} & \textbf{Unit}\\
      \hline
      \textbf{Dynamics:}\\
      Equivalent Isotropic Energy  &  $E_\mathrm{iso}$  & $10^{53}$ & erg \\
      CSM number density  & $n_0$   & $1$   & $\mathrm{cm}^{-3}$\\
      Initial peak Lorentz factor  &  $\Gamma_{\mathrm{peak},0}$ & $100$     &       \\
      Corresponding initial time & $t_0$ & $4.36 \times 10^6$ & s\\
      Jet half-opening angle    &  $\theta_\mathrm{jet}$  & 0.1 & rad \\
      Temperature of CSM ($p/\rho c^2)$ &  $\eta$ & $10^{-5}$ & \\
      \\
      \textbf{Micro-physics:} & & slow / fast &\\
      Fraction of accelerated e$^{-}$ &  $\xi_N$ & $1$ / $0.1$ & \\
      Electron energy &  $\epsilon_e$ & $0.1$ / $0.1$ & \\
      Magnetic energy &  $\epsilon_B$ & $0.01$ / $0.1$ & \\
      \hline

    \end{tabular}
    \label{tab:initial_parameters}
  \end{table}

  \subsection{Grid parameters}
  \label{sub:synth_lc_params}

  The grid contains $N_\theta = 300$ tracks in the interval $\theta \in [0, \pi/2]$. Initially we set the grid radial width as a function of initial shock Lorentz factor $\Gamma_s$ such that the radial bounds are equal to $r_{\mathrm{min},0} = r_s - 50 / \Gamma_s^2$ and $r_{\mathrm{max},0} = r_s + 50 / \Gamma_s^2$, where $r_s$ is the initial shock position as given by the BM solution. Each track contains $N_{r,0} = 5000$ cells of equal $\mathrm{d}r$ initially (except for $r>r_s$ where $dr$ is multiplied by a factor 10 only in the initial grid to make sure the blast-wave does not outrun the moving outer boundary in the first few time-steps) so as to resolve the blast wave correctly and ensure the energy contained in the grid is as close to the expected $E_\mathrm{iso} / (1 - \cos{\theta_\mathrm{jet}}))$ as possible. The number of cells per track quickly decreases due to the AMR criteria described in the next paragraph. During the simulation, we move the outer boundary such that the shock front on the jet axis is always located at $0.9 r_\mathrm{max}$. New cells are simply created from this moving boundary by the AMR methods implemented in the code. All boundaries apart from the outer one are reflective to ensure energy conservation in the grid (inner boundary), model potential interaction with the counter-jet (boundary at $\theta = \pi/2$), and comply with axisymmetry (boundary at $\theta = 0$).

  The resolution is determined as follows. The radial resolution is governed for each cell by a special "re-gridding score" $S_\mathrm{regrid} = \hat{a} \times \Gamma^{3/2}$ where $\hat{a}$ is the modified aspect ratio described in section \ref{sub:2d_simulation} and $\Gamma$ the fluid Lorentz factor at the cell location. $S_\mathrm{regrid}$ increases with Lorentz factor such that smaller aspect ratios are allowed for highly relativistic cells. This ensures that the blast wave is better resolved at the shock position where the velocity is the highest. We ran simulations with varying exponent values and found $\Gamma^{3/2}$ to give satisfying convergence of the light-curves. We also multiply $S_\mathrm{regrid}$ by a factor 10 in the few cells around the onset of the forward shock to further increase the resolution there. $S_\mathrm{regrid}$ is allowed to vary in the interval $[0.1, 3]$ leading to aspect ratios ranging from $10^{-4}$ to 3 in practice. The re-gridding mode is set to "runaway". In order to improve the resolution close to the jet axis, we use a variable-size track width where the position of interface $j-1/2$ between tracks $j-1$ and $j$ is set by:
  \begin{align}
    \theta_{j-1/2} = \frac{\pi}{2} \left(0.3\frac{j}{N_\theta} + 0.7\left(\frac{j}{N_\theta}\right)^3\right),
  \end{align}
  where $N_\theta$ is the number of tracks. This means that the radial resolution is also higher closer to the jet axis. The number of cells on track 0 varies from $\sim$450 to $\sim$1500 throughout the simulation. The dynamical simulations run in 7 hours on 384 Marvell ThunderX2 ARM cores distributed over 12 nodes.

  The synchrotron emission is computed following the method described in section 
  \ref{sub:local_synchrotron_cooling} with a small modification. The the larger size of the cells ahead of the blast wave leads to shock diffusion over the same scale as that of the hot region, and thus a wide region ahead of the blast wave is marked as "shocked" by the shock detector. We decide to turn off the emissivity in all the cells neighbouring a detected shock, and only sum over the emission from the cells in the process of cooling. This does not hinder the ability of the computed light curves to converge since the resolution directly downstream of the shocks is sufficient to allow for this approximation. We check that the light curves obtained are indeed converged by running simulations with varying radial and transverse resolutions (see sec. \ref{sub:synth_lc_results}, light curve with 600 tracks).

  \subsection{Results}
  \label{sub:synth_lc_results}

  \begin{figure*}
  \includegraphics[width=1.02\textwidth]{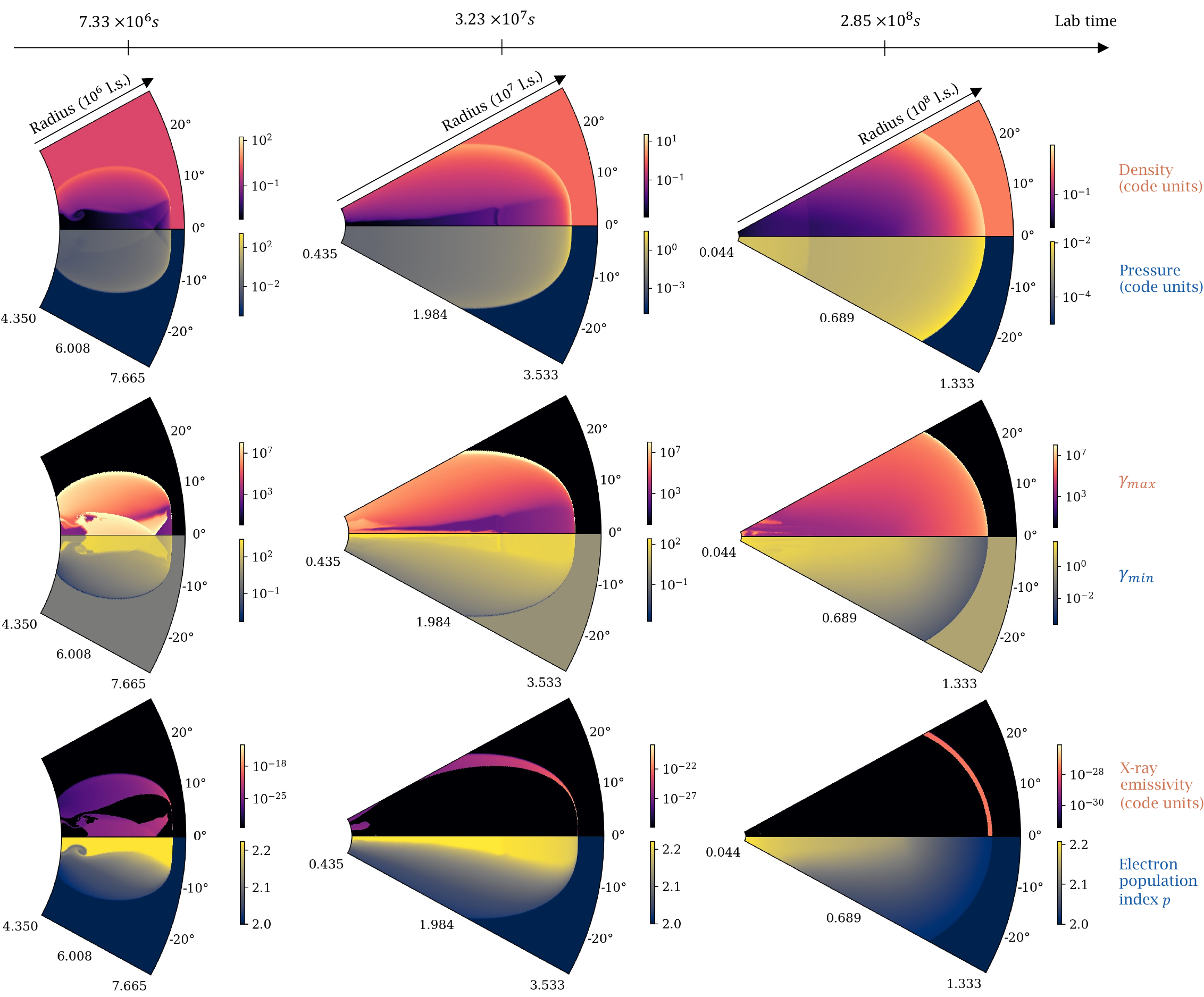}
  \caption{Snapshots in the slow-cooling case before spreading (left column), at $t_s$ (center column), and at the end of the simulation (right column). The snapshots are truncated at $\theta=0.5$ for better readability. The bottom line shows the x-ray emissivity (positive angles) as seen from an on-axis static observer. Since we use a log color scale, we floor the emissivity to the lowest measured non-zero value in the grid. This quantity allows us to map the emission sites in the jets and verify that the contribution to x-ray from internal re-collimation shocks early in the evolution is negligible in our simulation as the density downstream of these shocks is very low compared to the FS.}
  \label{fig:BoxFit_snapshots}
  \end{figure*}

  \begin{figure}
    \centering
    \includegraphics[width=0.45\textwidth]{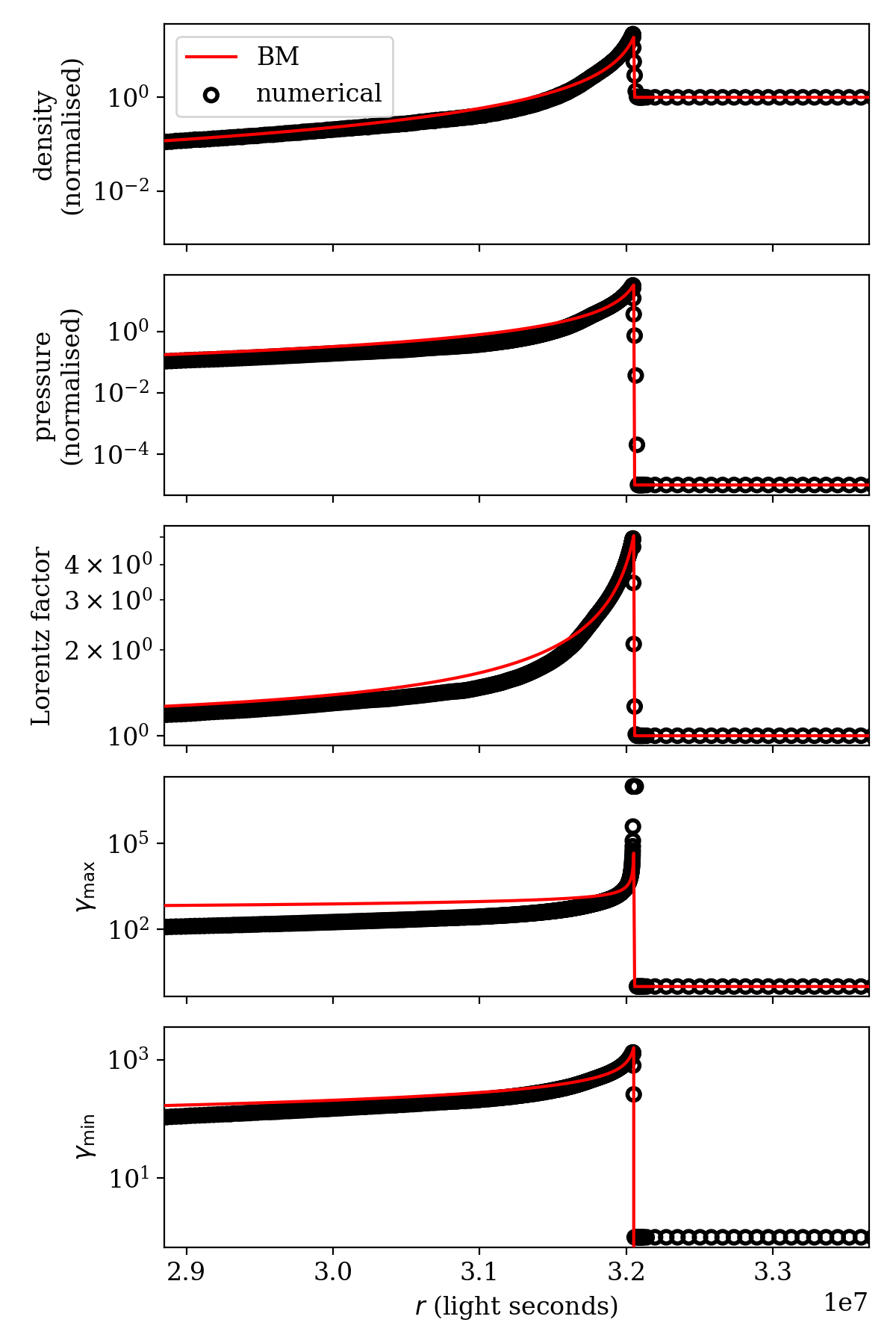}
    \caption{Radial profile at $t_s$ in the fast-cooling case for $\theta=0$. The jet still follows the BM solution at this stage. Density and pressure are normalised by $m_p n_0$ and $m_p \eta n_0 c^2$, respectively.}
    \label{fig:BoxfitProfile}
  \end{figure}

   \begin{figure}
      \centering
      \includegraphics[width=0.5\textwidth]{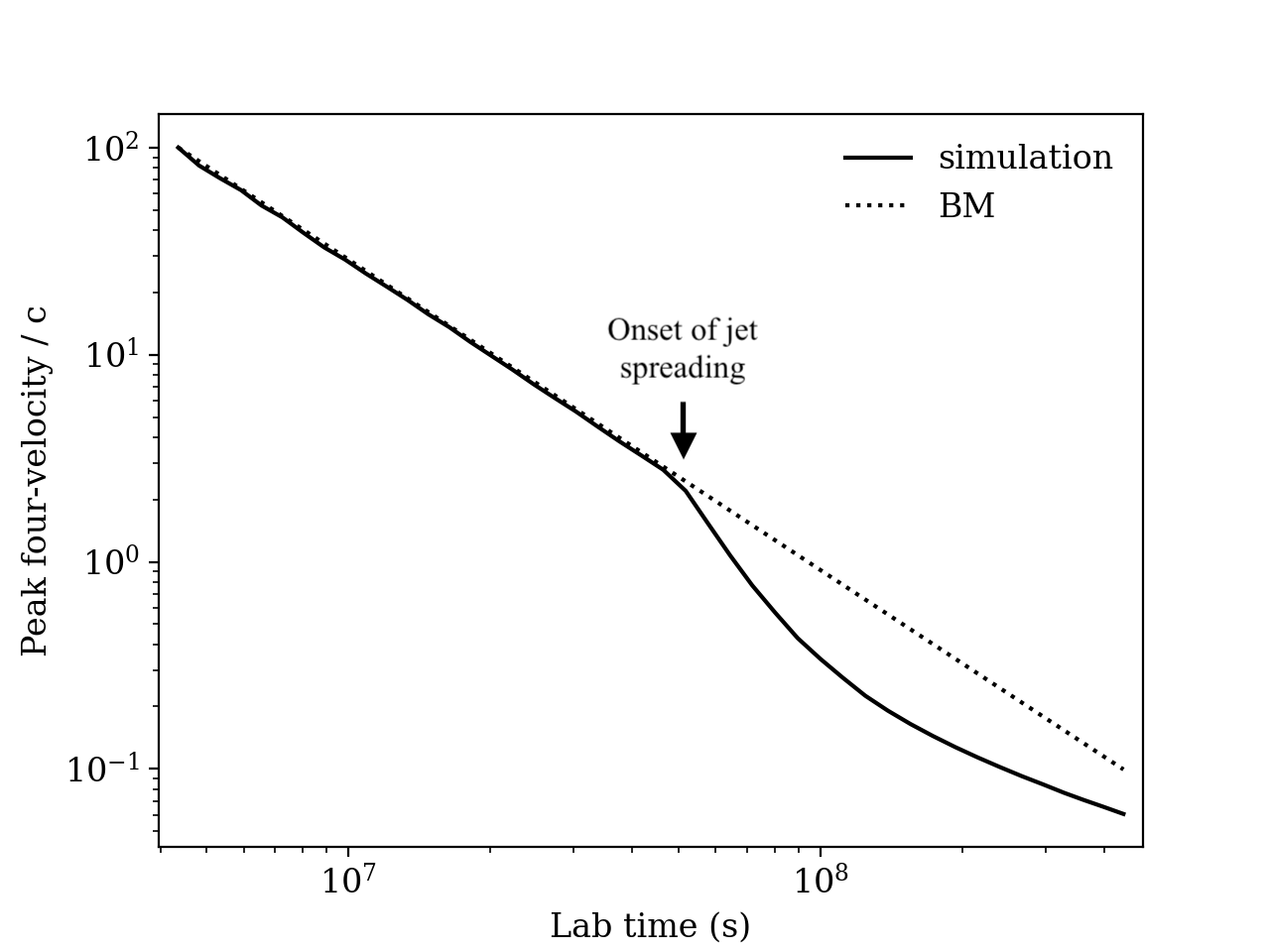}
      \caption{Four-velocity $\Gamma \beta$ directly downstream of the shock. We obtain this value by simply measuring the maximum radial velocity in the simulation domain. Our dynamical simulation very accurately follows the BM solution $\Gamma \beta \propto t^{-3/2}$ before jet spreading at $t_s = 3.22 \times 10^{7}$s.}
      \label{fig:fvelPeakvtime}
  \end{figure}

   \begin{figure}
      \centering
      \includegraphics[width=0.5\textwidth]{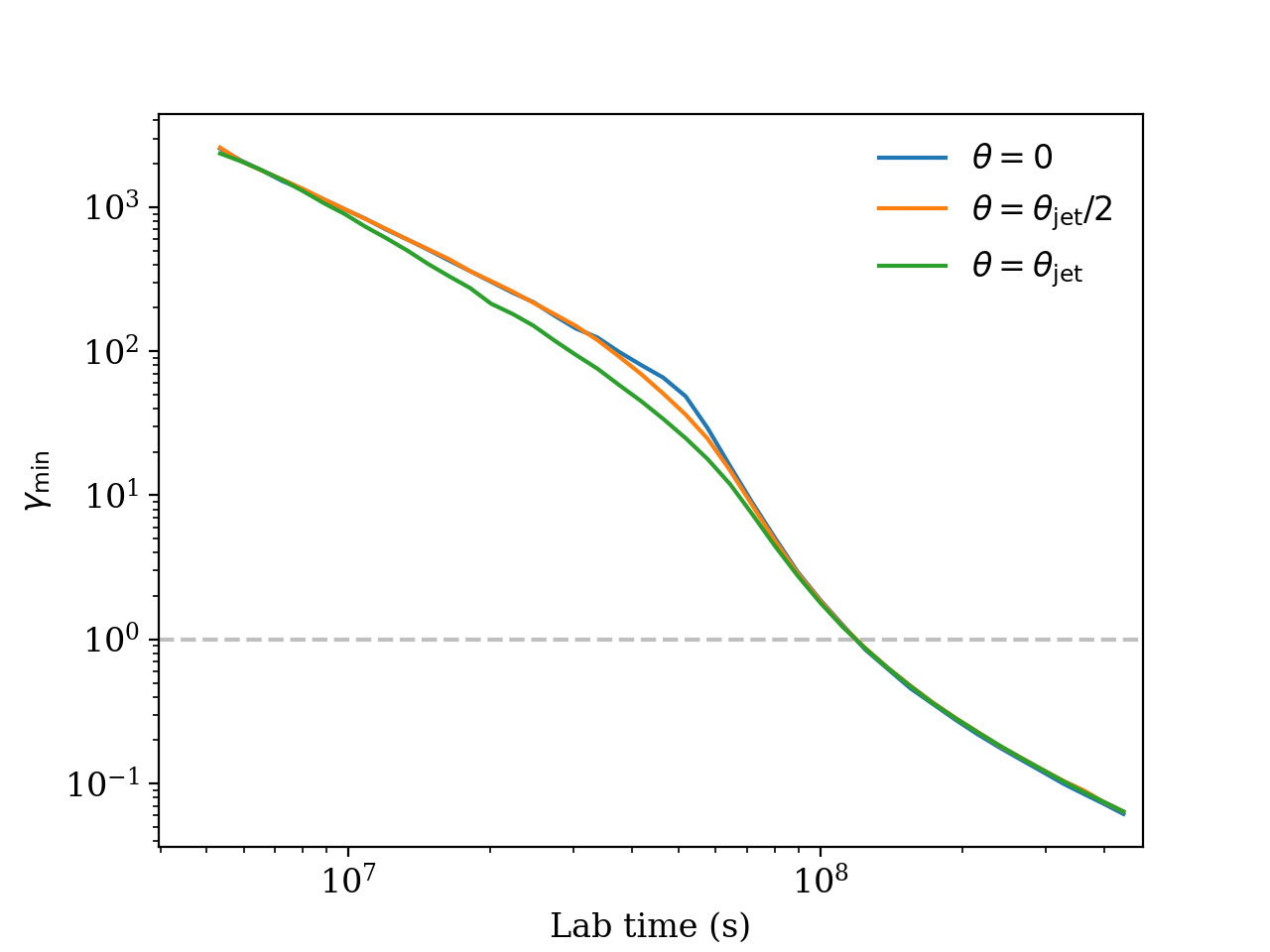}
      \caption{Slow-cooling case, fixed spectral index $p$. Lower bound $\gamma_\mathrm{min}$ of the accelerated electron population directly downstream of the shock for different directions (jet axis $\theta=0$, jet half-opening angle $\theta = \theta_\mathrm{jet}$, and halfway $\theta = \theta_\mathrm{jet}/2$). As expected $\gamma_\mathrm{min}$ decreases faster from the beginning due to jet edge effects. All evolutions rejoin when the jet enters the spherical expansion phase in the Newtonian regime. Eventually we encounter the known issue with describing the accelerated particles as a power-law in energy as $\gamma_\mathrm{min}$ moves below 1 (dashed gray line) at very late times.}
      \label{fig:gminPeakvtime}
  \end{figure}

   \begin{figure}
      \centering
      \includegraphics[width=0.5\textwidth]{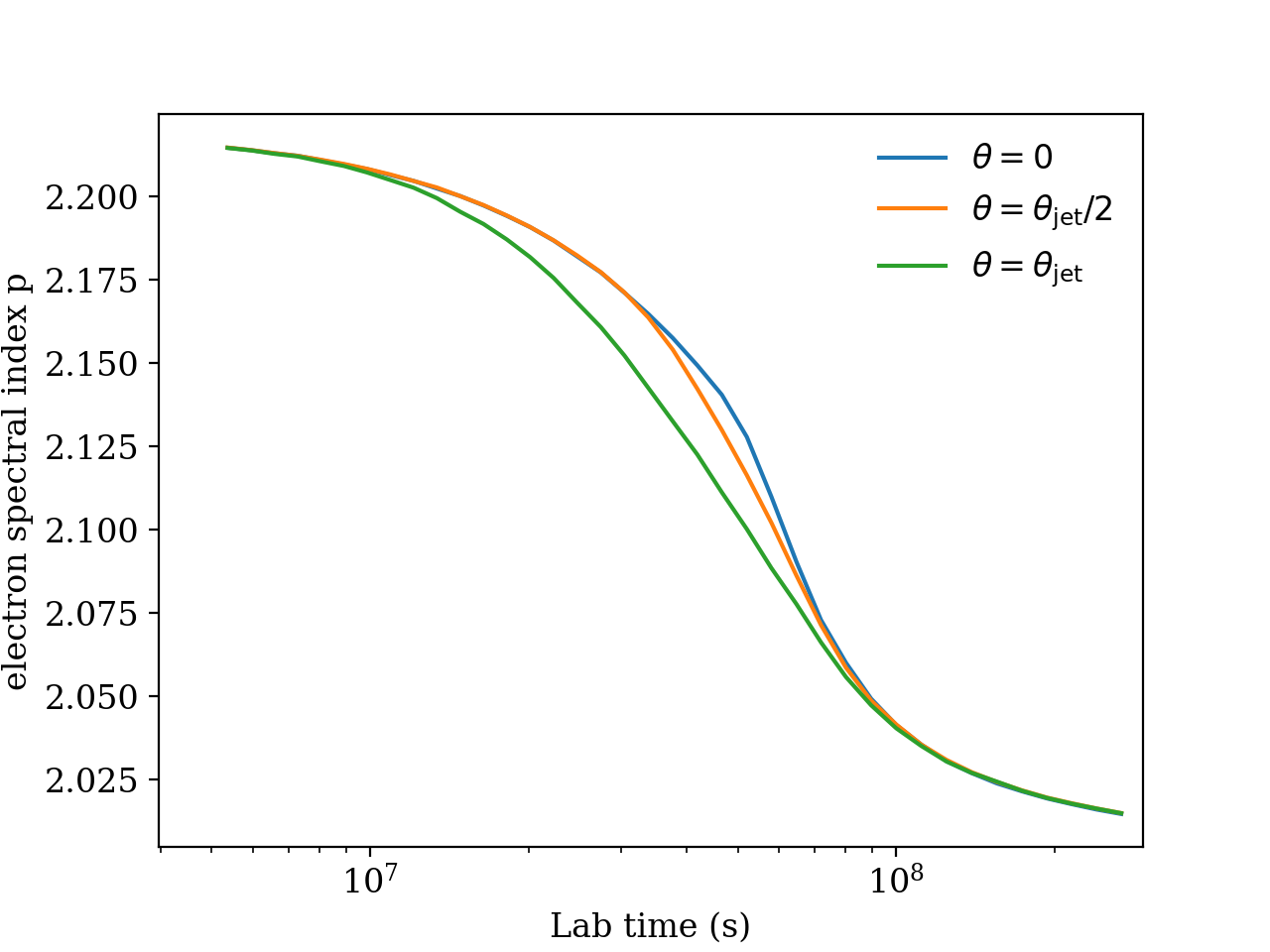}
      \caption{Slow-cooling case, variable spectral index $p$. Spectral index $p$ directly downstream of the shock for different directions (jet axis $\theta=0$, jet half-opening angle $\theta = \theta_\mathrm{jet}$, and halfway $\theta = \theta_\mathrm{jet}/2$). $p$ undergose a fast transition during jet spreading. This explains why the light-curves and spectra are identical in the variable and fixed $p$ cases in our figures.}
      \label{fig:pPeakvtime}
  \end{figure}

  \begin{figure}
    \centering
    \includegraphics[width=0.5\textwidth]{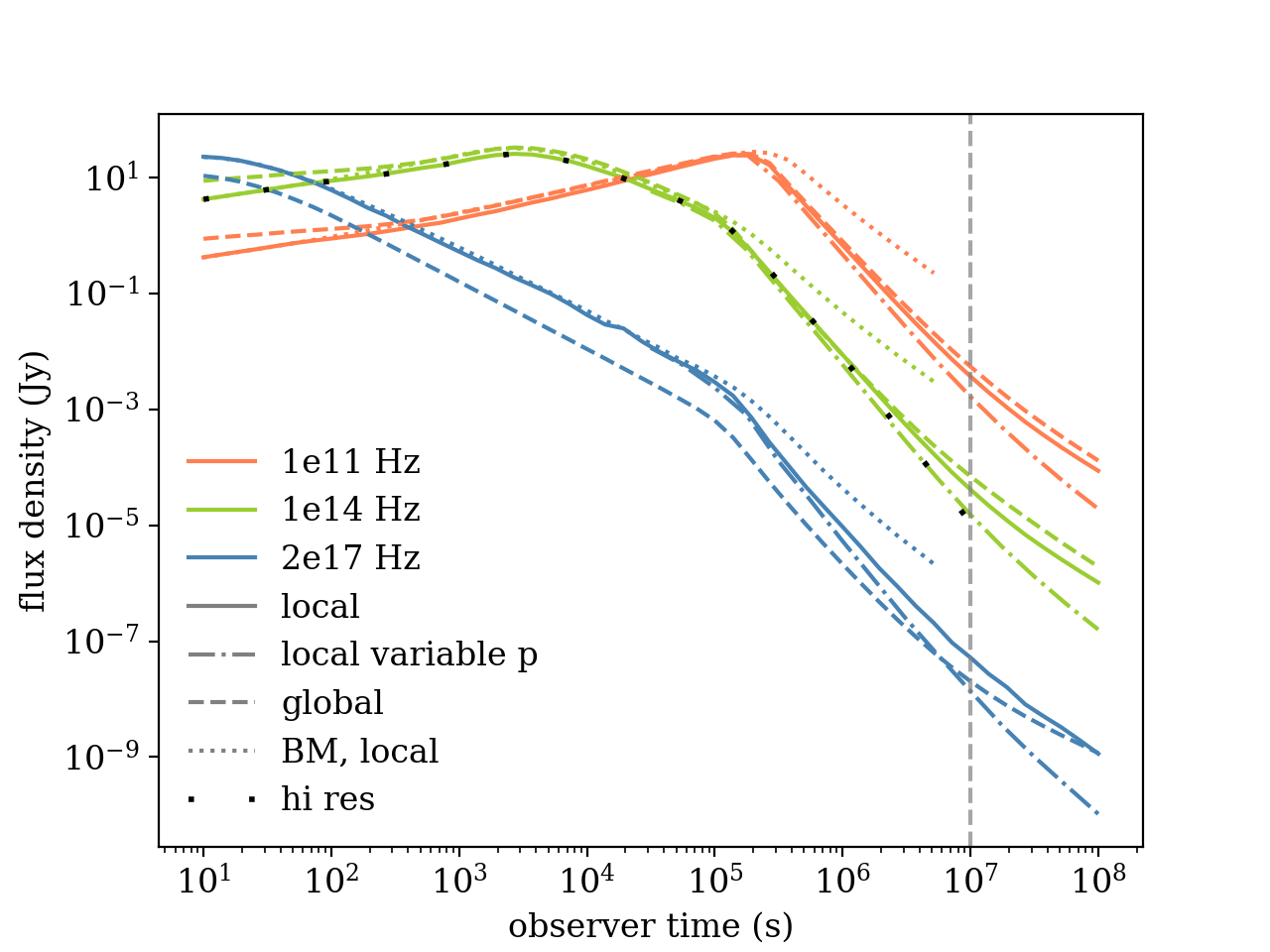}
    \caption{Light curves in the slow-cooling case in far IR (orange), optical (green) and x-ray (blue). The BM solution is not computed after $10^7$s as is is outside its region of validity then. As expected, it diverges from the true solution after the jet break at $t_\mathrm{obs}\sim 10^5$s. The difference between local and global cooling at early times for optical and far IR is due to the fact that the system is still fast-cooling then. As it transition to slow-cooling the light-curves rejoin. At later times, the effect of local cooling starts to show on $\gamma_\mathrm{min}$ that decreases faster than in the global cooling case, leading to lower fluxes above the injection break. After $t_\mathrm{obs}\sim 10^7$s (dashed gray vertical line), some of the contributing regions of the blast-wave see $\gamma_\mathrm{min}<1$ in the variable $p$ case and we caution against interpretation of the corresponding dot-dashed curves at later times. The loosely dotted black line shows the optical light-curve for a local, variable $p$ run at higher transverse resolution (600 tracks). Other frequencies are not shown for readability but we confirmed convergence there too.}
    \label{fig:slow_cooling_lcs}
    \includegraphics[width=0.5\textwidth]{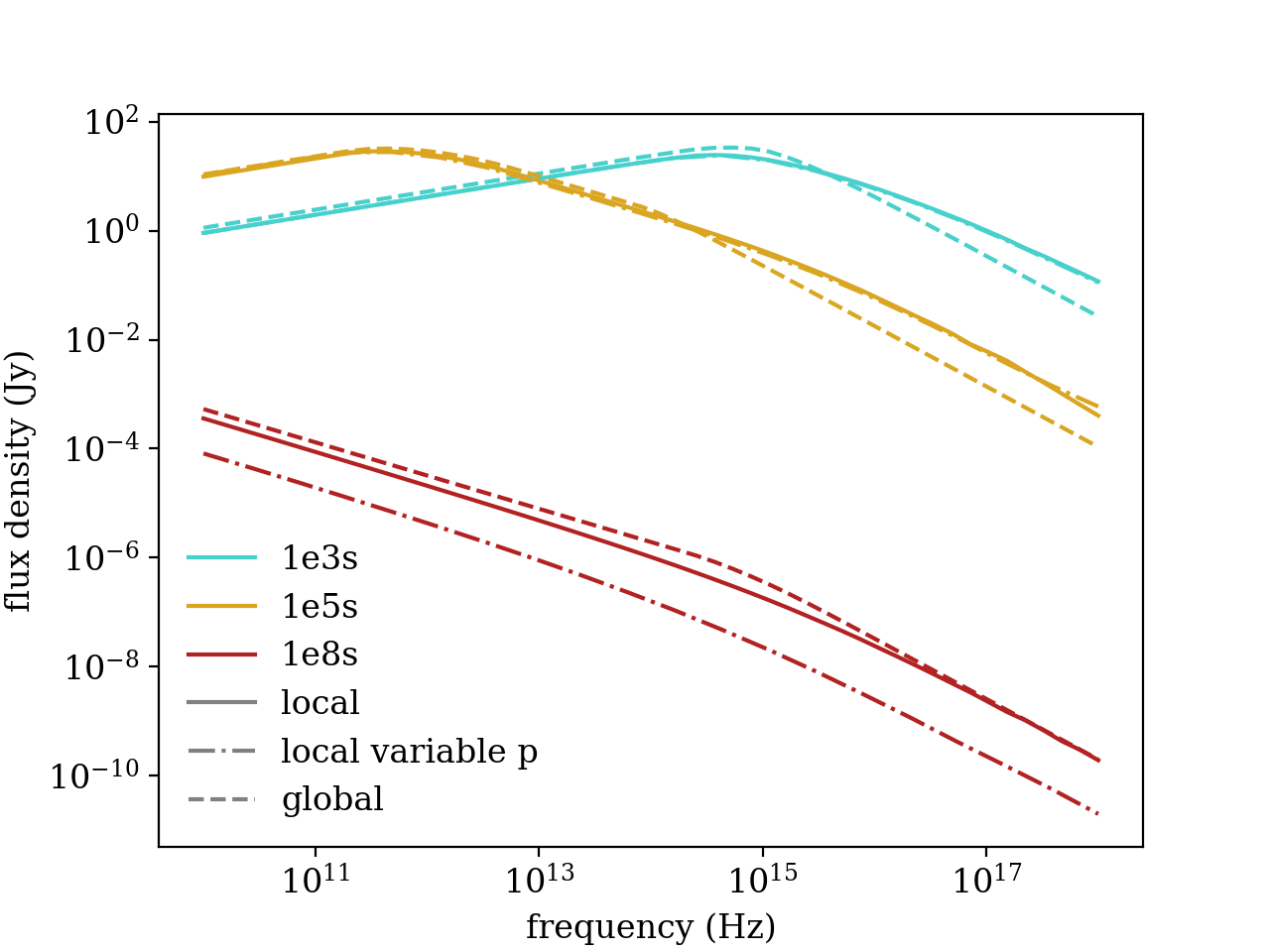}
    \caption{Slow-cooling spectra. Line styles match those of fig \ref{fig:slow_cooling_lcs}. Since this is slow cooling, the cooling break is easily identified as the right-most break in each spectrum. The difference in the cooling break position is responsible for the difference in flux density at high frequencies.}
    \label{fig:slow_cooling_spectra}
  \end{figure}

  \begin{figure}
    \centering 
    \includegraphics[width=0.5\textwidth]{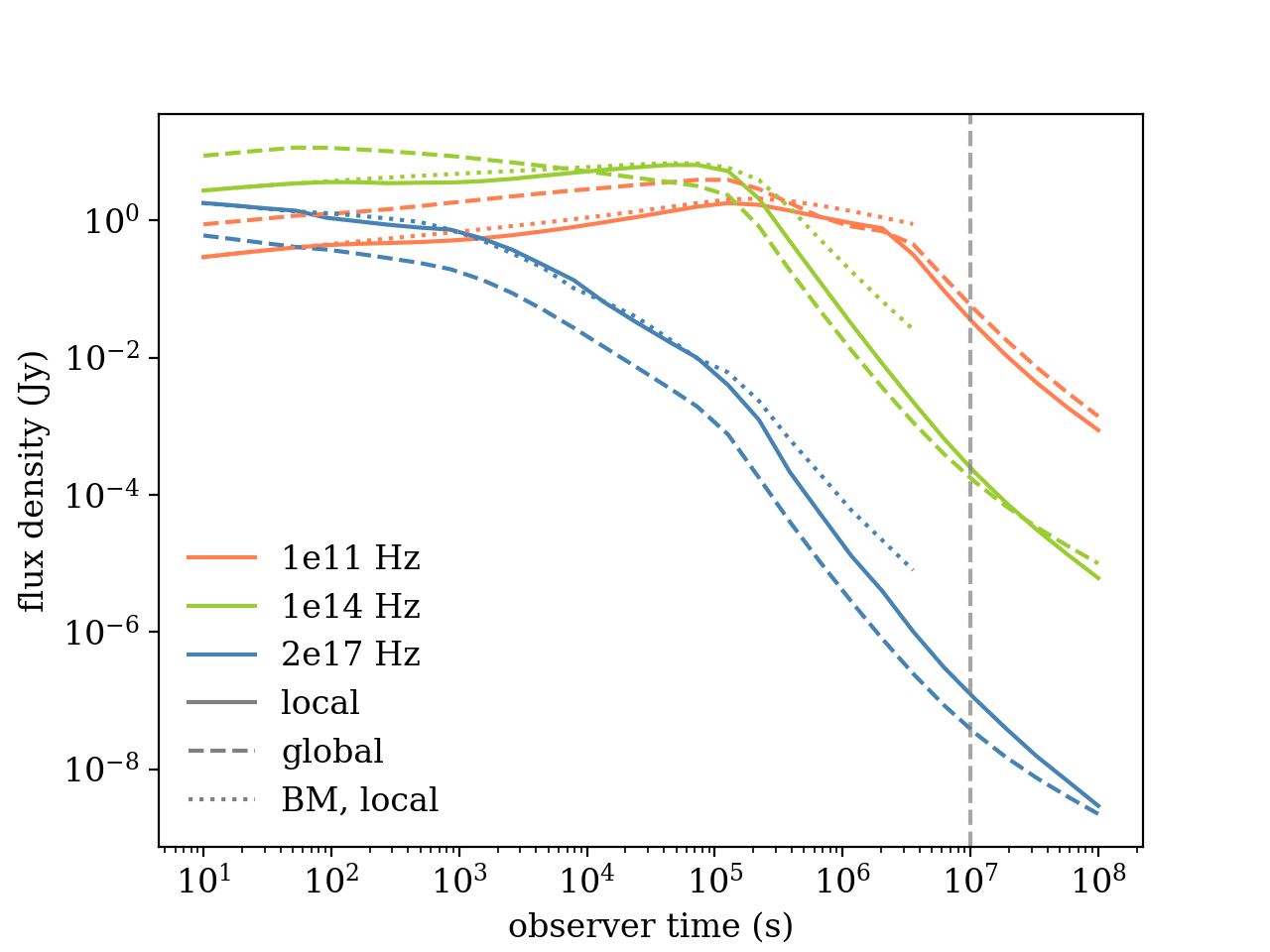}
    \caption{Same as figure \ref{fig:slow_cooling_lcs} but in the fast-cooling case. We do not show the variable $p$ light curve in this case which is discussed instead in the slow-cooling case. As in slow cooling all curves exhibit a jet break at $t_\mathrm{obs}\sim 10^5$s. Just like in slow-cooling, the x-ray light-curve shows a factor $\sim5$ difference pre-jet break as this frequency is placed above the cooling break throughout the whole evolution.}
    \label{fig:fast_cooling_lcs}
    \includegraphics[width=0.5\textwidth]{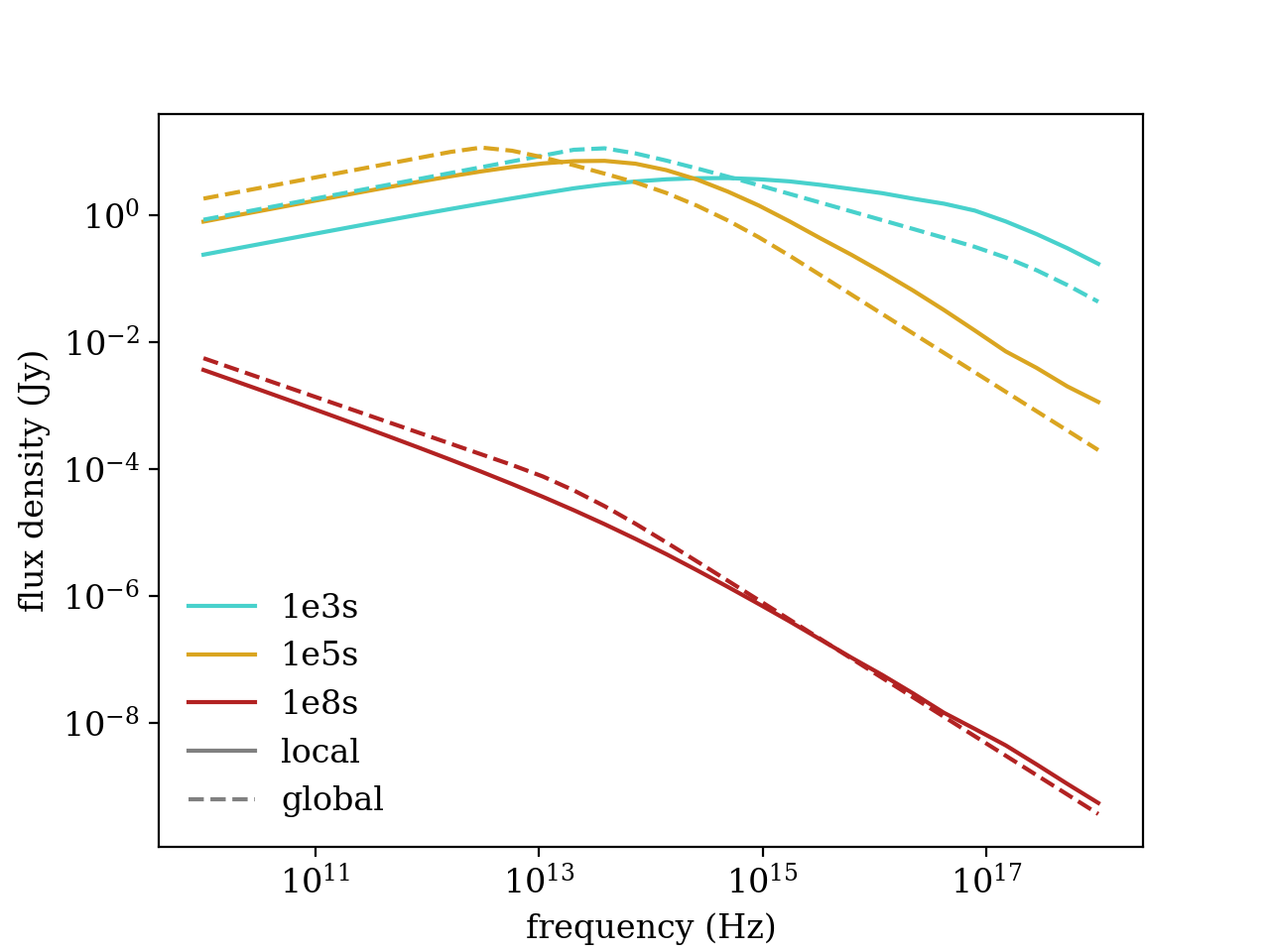}
    \caption{Fast-cooling spectra. Line styles match those of fig \ref{fig:slow_cooling_lcs}.}
    \label{fig:fast_cooling_spectra}
  \end{figure}

  Snapshots in the slow-cooling case at $1.6\times t_0$, spreading time $t_s$ and close to final time $t \lesssim t_f$ are reported in figure \ref{fig:BoxFit_snapshots}. For better readability, these snapshots are truncated at $\theta=0.5$~rad. We observe the expected behavior where the first stages of the evolution follow the BM solution as the jet lacks transverse causal connection. The jet eventually starts spreading at:
  \begin{align}
    t_s = \left( \frac{17E_\mathrm{iso} \theta_\mathrm{jet}^2}
                        {4\pi n_0 m_p c^5} \right) ^{1/3}
    = 3.22 \times 10^7\mathrm{s},
  \end{align}
  where it transitions to the spherical evolution stage in which we choose to stop the simulation at $t_f$. In figure~\ref{fig:BoxfitProfile} we show that at $t_s$ the blast wave still follows the BM solution. this figure also highlights the very high resolution needed to properly resolve the hot region where $\gamma_\mathrm{max}$ quickly decreases downstream of the shock. In figure~\ref{fig:fvelPeakvtime} we show that our simulation accurately follows the expected BM evolution until jet spreading. The peak four-velocity directly downstream of the shock $u_\mathrm{Peak}$ follows the expected evolution with lab time $u_\mathrm{Peak} \propto t^{-3/2}$ very closely until the onset of transverse motion (jet spreading). Figure \ref{fig:gminPeakvtime} shows the peak value for $\gamma_\mathrm{min}$ (directly downstream of the shock) as a function of lab time in three different directions. Here too, we recover the expected evolution pre-jet break except near the outer edge of the jet where interaction with the CSM influences $\gamma_\mathrm{min}$ from the start. This figure highlights the transition to the spherical expansion phase when $\gamma_\mathrm{min}$ adopts the same evolution in all plotted directions for $t > 10^{8}$s. At these late times, with our description of the particle population, we enter the regime in which $\gamma_\mathrm{min} < 1$. When computing the emissivity from these regions, we apply the prescription in the deep newtoninan phase from section \ref{sub:deep_newtonian_phase}, where $\gamma_\mathrm{min}$ is floored to 1. We report in the figures the limit observer time $t_\mathrm{lim} \sim 10^7$s above which regions of the fluid with such low values of $\gamma_\mathrm{min}$ start contributing to the light-curves. Figure \ref{fig:pPeakvtime} shows the evolution the spectral index $p$ directly downstream of the shock with lab time, in the case where we allow it to vary. We can notice it does not evolve significantly before the jet break here and undergoes a sharp decrease as the jet decelerates post-jet break. This is easily explained by the direct dependency of $p$ on the upstream fluid velocity.

  The synthetic light curves and spectra are reported in figures \ref{fig:slow_cooling_lcs}-\ref{fig:slow_cooling_spectra} (slow-cooling) and \ref{fig:fast_cooling_lcs}-\ref{fig:fast_cooling_spectra} (fast-cooling). All light-curves exhibit a jet break at $t_\mathrm{obs} = t_\mathrm{break} \sim 10^5$s corresponding to the spreading time $t_s$ in the lab frame. In both cases, we observe pre-jet break a very good match with the expected BM flux computed with local cooling \citep{Granot2001}, which confirms the validity of our approach. Post-jet break, the BM solution diverges from the true solution as it does not take into account jet spreading and the associated deceleration. As expected, we observe a steeper decrease post-jet break when accounting for jet spreading and deceleration in our numerical solution. 

  We can now compare the light curves and spectra to those obtained with the global cooling approximation. Let us first consider the slow-cooling case. We can observe a discrepancy of factor $\sim5$ in flux levels between the global cooling and local cooling approaches after the cooling break. This is explained by the difference in the position of the cooling break in the spectrum, which has a direct influence on the overall flux level in the light-curves at higher frequencies. Now considering the fast-cooling case, we see that the change in cooling break frequency influences all parts of the spectrum as $\gamma_\mathrm{min}$ is now also subject to cooling. The light curves flux level at frequencies below $\nu_m$ is thus also affected in this case. These strong differences highlight the clear need to include local cooling in the modeling tools currently used by the community.

\begin{figure}
    \centering
    \includegraphics[width=0.5\textwidth]{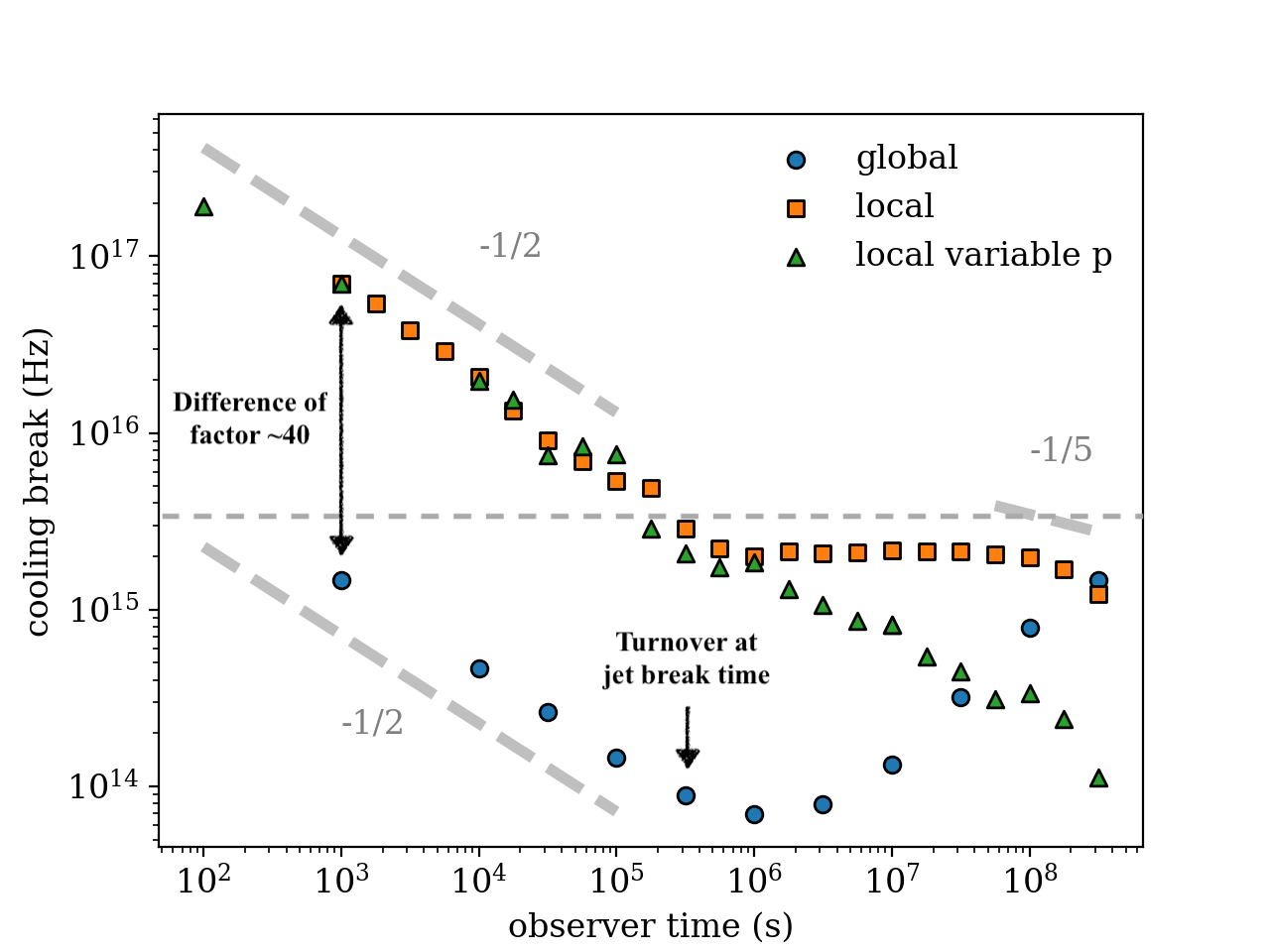}
    \caption{Evolution of the cooling break position with observer time. The dashed segments represent the asymptotic slopes expected in the relativistic (-1/2) and Newtonian (-1/5) limits. The global cooling results match those from \citet{VanEerten2012a} and \citet{VanEerten2013a} with a temporary sharp increase of $\nu_c$. The local cooling approach displays a very different behavior with constant $\nu_c$ for at least two decades in observer time from the jet break onwards. The horizontal dashed gray line is the frequency used to plot the evolution of the spectral index in figure \ref{fig:spectralIndex}.}
    \label{fig:nuc_evol}
  \end{figure}

  Locally tracing the particle population offers several opportunities regarding the study of the evolution for the emission post-jet break. We can for the first time accurately capture the radiative transition from the ultra-relativistic BM solution to the Newtonian ST solution. In figure \ref{fig:nuc_evol} we report the evolution of the cooling break $\nu_c$ with observer time $t_\mathrm{obs}$. Firstly, we find that our simulations using the global cooling approach are in good agreement with previous trans-relativistic simulation works \citep{VanEerten2012a,VanEerten2013a}. We find the expected $-1/2$ slope in the ultra-relativistic limit and observe the same turnover at $t_\mathrm{break}$. Secondly, with local cooling, $\nu_c$ follows the same slope for $t_\mathrm{obs}<t_\mathrm{break}$, but offset by a factor $\sim 40$. This feature was expected as pointed out by previous works \citep{VanEerten2010a,Guidorzi2014}. Post-jet break, we observe a striking difference between the two approaches. While $\nu_{c,\mathrm{global}}$ sharply increases, $\nu_{c,\mathrm{local}}$ transitions to a plateau stage from $t_\mathrm{obs}\sim 3 \times 10^5$s to $t_\mathrm{obs}\sim 3 \times 10^7$s. As the jet transitions to the Newtonian phase, $\nu_c$ resumes decreasing. In figure \ref{fig:nuc_evol} we show the asymptotic -1/5 slope expected for this phase. Unfortunately, our simulations did not run long enough to allow us to model the radiation at later times yet and we cannot confirm at this stage that $\nu_c$ will settle on this asymptote from this simulation. In this particular setup the global and local calculations of $\nu_c$ meet up at at times. Investigating whether this phenomenon happens for all explosion parameters or has a physical explanation is left to future work.

\begin{figure}
    \centering
    \includegraphics[width=0.5\textwidth]{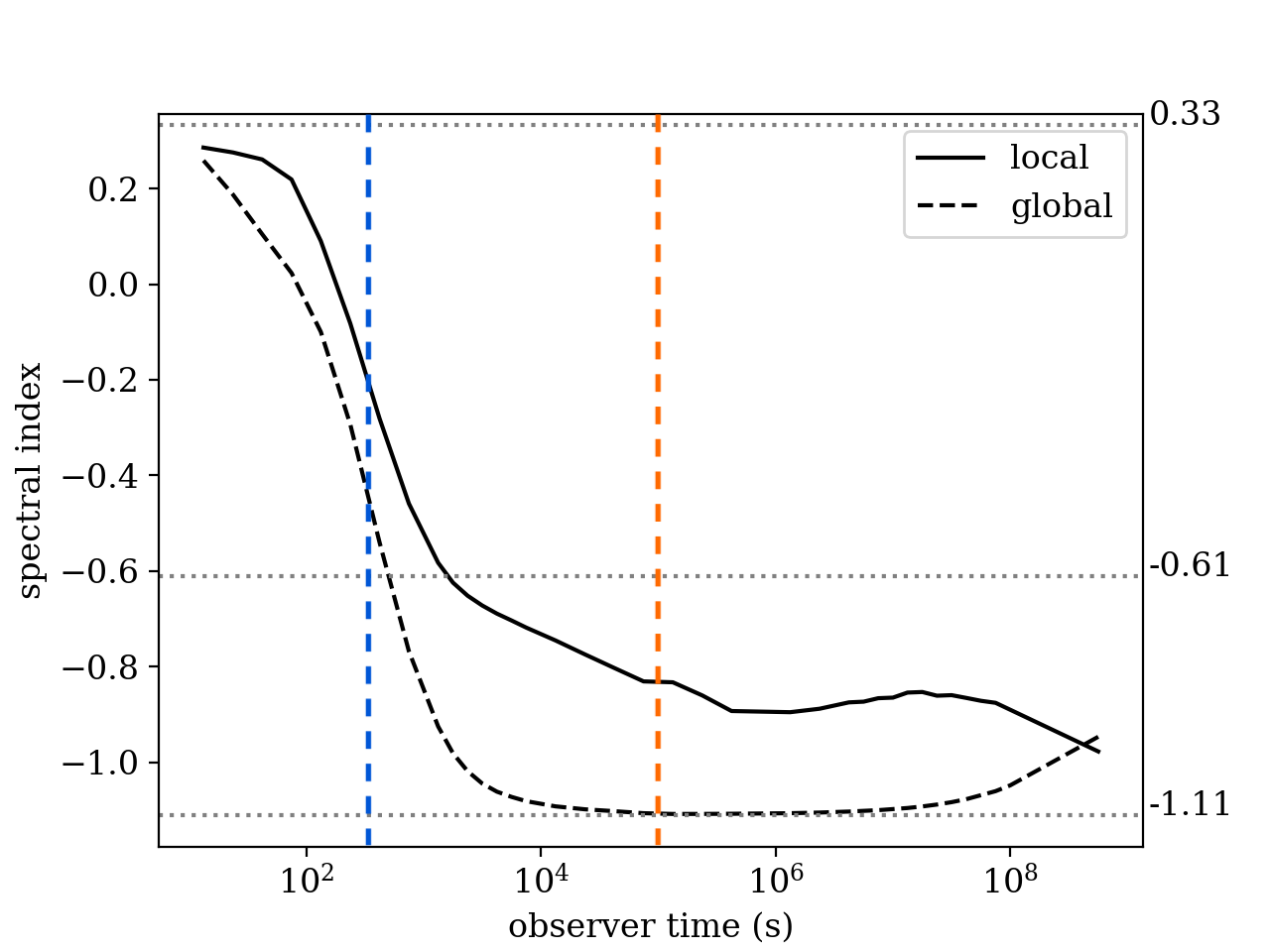}
    \caption{Evolution of the spectral index $\beta$ in the slow-cooling case at frequency $f = 3.16 \times 10^{15}$~Hz as a function of observer time. The local cooling calculation is done for a fixed value of $p$. The blue (orange) dashed line represents the approximate observer time at which the global (local) $\nu_c$ crosses our observed frequency. At $10$s, our $f$ sits below $\nu_m$ and $\nu_c$ for both cooling prescriptions, at the expected value $\beta = 1/3$. As we are just exiting the fast-cooling stage, $\nu_m$ and $\nu_c$ are very close together, and, in the global cooling case, $\beta$ does not have the time to settle on the expected -0.61 value when $\nu_m$ crosses $f$, and instead drops straight to -1.11, which is the value expected for $f>\nu_c$. In the slow-cooling case, $\nu_c$ sits at higher frequencies, and the cooling break in the spectrum is smoother, which leads to a slower transition towards -1.11. Additionally, as the evolution of $\nu_c$ flattens out post-jet break, our observing frequency $f$ never leaves the break region, and -1.11 is never reached.}
    \label{fig:spectralIndex}
  \end{figure}

  The difference in behaviour of $\nu_c$ between the local cooling and global cooling approaches is also visible in figure \ref{fig:spectralIndex} where we show the evolution of the spectral index $\beta$ at $3.16 \times 10^{15}$~Hz, computed across a decade in frequency, as a function of observer time. This frequency does not correspond to a meaningful observing instrument but is chosen since $\nu_c$ crosses it in both the global and local cooling cases. The cooling break transition occurs at later times for local cooling in comparison with global cooling, as expected from figure \ref{fig:nuc_evol}. The cooling break is also known to be very smooth \citep{Granot2001,VanEerten2009a,Uhm2014}. We recover this effect as we observe that $\beta$ transitions more slowly to the value expected above the jet break of -1.11. In our case, the smoothness is obtained from the sum of the contributions to the spectrum of all regions of the fluid. Since our local emission coefficient is a sharp broken power-law and includes a sharp cut-off above $\nu_\mathrm{max}$, we expect to actually be underestimating the smoothness of this transition.

\begin{figure}
    \centering
    \includegraphics[width=0.5\textwidth]{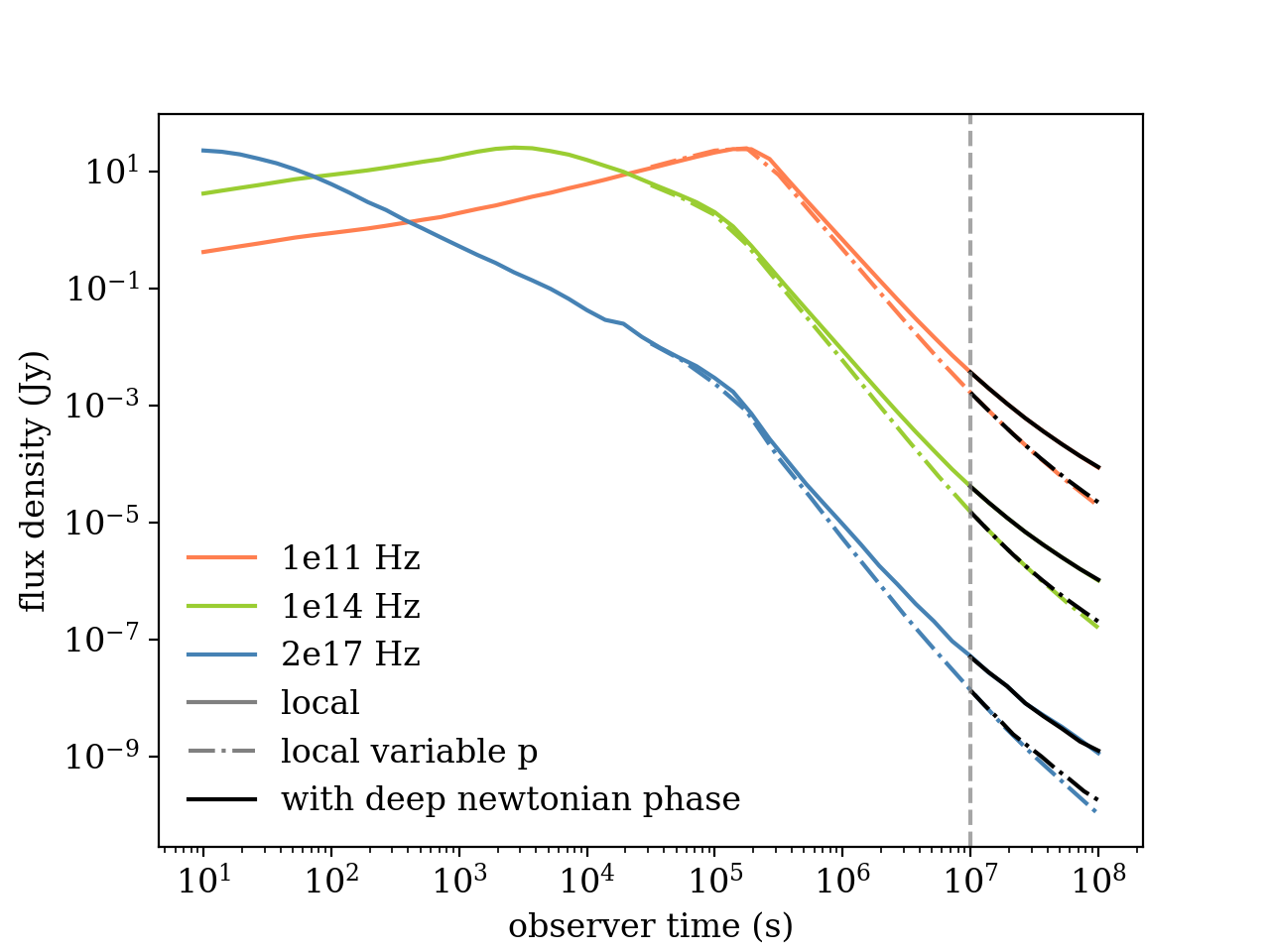}
    \caption{Effect of the inclusion of the calculation of the deep Newtonian phase on the resulting light curves in the slow-cooling case. In color are the light curves computed without flooring $\gamma_\mathrm{min}$ and computing the contribution only from the relativistic electrons. In black are the corrected light curves to include the deep Newtonian regime. The deep Newtonian phase is only shown past $10^7$s as $\gamma_\mathrm{min}>1$ before this time in the whole fluid. The effects of the deep Newtonian regime are marginal at the times simulated here, except in the case of the x-ray light curve for a variable spectral index p, where it contributes to a light flattening of the light curve at late times.}
    \label{fig:DNP}
  \end{figure}

  In all the figures describing the radiation from our the slow-cooling case, we also show the evolution of $\nu_c$ in the case where the spectral index $p$ varies following the approach described in section \ref{sub:local_synchrotron_cooling}. We observe further differences for $t>t_\mathrm{break}$ from the local cooling approach where it is kept to a fixed value $p=2.22$, where the flux decreases faster at all frequencies. The collapse in flux is directly linked with the sharper decrease of $\gamma_\mathrm{min}$ associated with decreasing spectral index after $t_s$, as is visible in the snapshots in figure \ref{fig:BoxFit_snapshots}. In this particular case, the influence of the deep Newtonian phase can be seen on the x-ray light curve as reported in figure \ref{fig:DNP}, in which we compare light curves calculated with, and without implementation of the Newtonian evolution. The flattening of the light curve at other wavelenghts, as well as for a fixed value of $p$, is marginal.

  The evolution of the cooling break position also changes and we observe in figure \ref{fig:nuc_evol} that the plateau at the jet break time present with a fixed spectral index disappears with a variable value for $p$, showing a steady decrease similar to the pre-jet break regime. This effect is visible in light curves and spectra and provides a potential avenue to investigate the theoretical processes involved in particle acceleration at shock fronts.

\section{Discussion}
\label{sec:discussion}


  In this work, we present a 2D relativistic hydrodynamics code that includes a local calculation of particle population evolution. While the use of a moving mesh offers significant improvements in terms of efficiency over fixed mesh approaches, the immediate downside of the local cooling approach is the necessity to run separate expensive dynamical simulations for each set of micro-physical parameters, increasing the number of runs necessary to explore the parameter space in comparison to global cooling approaches. 

  Regarding our jet simulations, we assume either sphericity or axi-symmetry and run 2D calculations. We do not expect 3D effects to strongly influence our synthetic light-curve calculations, however, the study of the Rayleigh-Taylor instabilities at the contact discontinuity in the afterglow would benefit from a 3D approach as this could have important consequences on the rate of propagation of the reverse shock in the ejecta. We also consider in these simulations the magnetisation to be small enough that it does not influence the general dynamics of the jet, which is expected at late stage of the evolution post-deceleration \citep[for a recent review, see e.g.][]{Granot2015}.

  The light curves and spectra presented in this work are all calculated for initially top-hat jets for an on-axis observer. Since GRBs are sources at cosmological distances, they are typically observed near on-axis or at least within the jet half-opening angle \citep{Ryan2015}, a situation for which angular structure of the jet has a negligible influence on the light-curve shape \citep{Rossi2002,Kumar2003,Ryan2019a}. These results can thus already be applied to the bulk of GRB afterglow observations. The influence of local cooling on structured jets observed off-axis will be the focus of a future study.


  In section \ref{sub:deep_newtonian_phase}, we explain how we approach the problem of values of $\gamma_\mathrm{min}$ reaching unphysical values below unity. While our approach produces accurate light curves in the deep Newtonian phase, it does not provide an accurate local accelerated electron distribution at low energies in the simulated dynamical snapshots, which makes the study of the interaction between the non-thermal and thermal components impossible at this stage. 
  Additionally, our model runs into limitations as $p$ approaches 2 (see eq. \ref{eq:DNP}). The emissivity becomes strongly dependent on the values of $\gamma_\mathrm{max}$ and $p$, which is either very hard to resolve numerically for the former, or poorly understood for the latter. For these reasons, a future version of \GAMMA will include a full local calculation of the distribution \emph{in momentum} of the non-thermal population.

  In section \ref{sub:radiative_flux_calculation}, we mention our simplified approach to the cut-off of local emissivity above $\nu_\mathrm{max}$. Since the slope above the cooling break in the observed spectrum is the result of this cut-off being placed at different frequencies in the observer frame depending on the emitting region of the blast-wave, a sharp cut-off will have a tendency to underestimate the flux received above $\nu_c$. However, this can be compensated by increased resolution in the emitting region, and we actually observed observe a very good agreement between our numerical approach and analytical solutions before the jet break. Our main results remain valid at later times too since what we observe is a significant increase in the observed flux with local cooling.

  In conclusion, this work provides a solution to the current pitfalls of numerical modeling of afterglow light-curves thanks to an improved numerical approach that accounts for the local variability of the emissivity in the fluid. The striking difference in cooling break behaviour across the jet break between local and global cooling approaches implies that it is not sufficient to merely apply a fixed correction factor to a global cooling approach in order to match the more physically realistic local cooling results. Nevertheless, the cooling break evolution curve remains completely scale invariant in the manner first described by \citet{VanEerten2012a} even across the trans-relativistic stage. 

  Recent discoveries associated with the multi-messenger detection of 170817 \citep{Abbott2017,Abbott2017b,Goldstein2017,Hallinan2017,Savchenko2017,Troja2017} have given new impetus for the development of more accurate numerical models. These are needed for us to be able understand the added complexity (jet structure, off-axis observer, kilonova contribution, interaction with a dynamical ejecta) from these ongoing observations observations \citep[e.g][]{Troja2019,Hajela2019a,Troja2020}. While the simulations presented here are a textbook case of top-hat on-axis GRB afterglow evolution, \GAMMA~now provides the basis for the implementation of more complex micro-physical descriptions for the emission and GRB dynamics which will be explored in future works.

\section*{Acknowledgements}

We thank the anonymous referee for a constructive report.
This work used the Isambard 2 UK National Tier-2 HPC Service (\url{http://gw4.ac.uk/isambard/}) operated by GW4 and the UK Met Office, and funded by EPSRC (EP/T022078/1).
This research made use of the Balena High Performance Computing (HPC) Service at the University of Bath. H. J. van Eerten acknowledges partial support by the European Union Horizon 2020 Programme under the AHEAD2020 project (grant agreement number 871158).

\section*{Data Availability}
 
The data underlying this article will be shared on reasonable request to the corresponding author.



\bibliographystyle{mnras}
\bibliography{Bath_PhD} 







\bsp  
\label{lastpage}
\end{document}